# Genesis of a Horizontal Electric Field within the Lipid Bilayer: A Bilayer-Embedded Actuation Platform

*Maki Komiya, Madoka Sato, Teng Ma, Hironori Kageyama, Tatsuya Nomoto, Takahisa Maki, Masayuki Iwamoto, Miyu Terashima, Daiki Ando, Takaya Watanabe, Yoshikazu Shimada, Daisuke Tadaki, Hideaki Yamamoto, Yuzuru Tozawa, Ryugo Tero, Albert Martí, Jordi Madrenas, Shigeru Kubota, Fumihiko Hirose, Michio Niwano, Shigetoshi Oiki*, Ayumi Hirano-Iwata**

M. Komiya, M. Sato, H. Kageyama, T. Nomoto, M. Terashima, D. Ando, T. Watanabe, Y. Shimada, D. Tadaki, H. Yamamoto, M. Niwano, A. Hirano-Iwata
Laboratory for Nanoelectronics and Spintronics, Research Institute of Electrical Communication (RIEC), Tohoku University, Sendai, Miyagi 980-8577, Japan.
E-mail: ayumi.hirano.a5@tohoku.ac.jp

M. Sato, H. Kageyama, T. Nomoto, T. Watanabe, A. Hirano-Iwata
Graduate School of Biomedical Engineering, Tohoku University, Sendai, Miyagi 980-8577, Japan.

T. Ma
School of Mechatronical Engineering, Beijing Institute of Technology, Beijing 100081, China

T. Ma, H. Yamamoto, M. Niwano, A. Hirano-Iwata
Advanced Institute for Materials Research (WPI-AIMR), Tohoku University, Sendai, Miyagi 980-8577, Japan.

T. Maki, M. Iwamoto
Department of Molecular Neuroscience, Faculty of Medical Sciences, University of Fukui, Yoshida, Fukui 910-1193, Japan.

T. Maki, M. Iwamoto
Life Science Innovation Center, University of Fukui, Fukui 910-1193, Japan.

D. Ando, Y. Shimada, A. Hirano-Iwata
Graduate School of Engineering, Tohoku University, Sendai, Miyagi 980-8577, Japan.

Y. Tozawa
Graduate School of Science and Engineering, Saitama University, Saitama, Saitama 338-8570, Japan.

R. Tero
Department of Applied Chemistry and Life Science, Toyohashi University of Technology, Toyohashi, Aichi 441-8580, Japan.

A. Martí, J. Madrenas
Department of Electronics Engineering, Universitat Politècnica de Catalunya, Barcelona, Catalunya, Spain.

S. Kubota, F. Hirose
Graduate School of Science and Engineering, Yamagata University, Yonezawa, Yamagata 992-8510, Japan.

S. Oiki
Biomedical Imaging Research Center, University of Fukui, Yoshida, Fukui 910-1193, Japan.
E-mail: oiki@u-fukui.ac.jp

Funding:
Japan Society for the Promotion of Science (JSPS) KAKENHI 23H00251 (AH-I), 26H02216 (AH-I), 24K17593 (MK), 20H00497 (SO), 24K21270 (SO), 24K01988 (MI), 24K21943 (MI), 25K10165 (TM)
Grant-in-Aid for JSPS Fellows 22J13311 (MS), 24KJ0348 (HK)
Japan Science and Technology Agency (JST) CREST JPMJCR14F3 (AH-I)
MEXT Grant-in-Aid for Transformative Research Areas (A) "Multicellular Neurobiocomputing" 24H02329 (HY), 24H02332 (HY), 24H02334 (AH-I)
Takeda Science Foundation Bioscience Research Grant (AH-I)
WISE Program for AI Electronics, Tohoku University (MS, HK)

Advanced Graduate Program for Future Medicine and Health Care, Tohoku University (TN)

Cooperative Research Project of the Research Institute of Electrical Communication, Tohoku University (RT, AH-I)

**Abstract:**

The electric field of biological membranes has long been treated as a one-dimensional quantity, defined solely by the component normal to the bilayer ($\boldsymbol{E}_{\mathrm{VERT}}$). Here, we present a bioelectronic platform that enables controlled generation of a horizontal electric field within the hydrophobic core of a lipid bilayer ($\boldsymbol{E}_{\mathrm{HORZ}}$). The device incorporates micrometer-scale electrodes embedded within the bilayer torus, allowing sustained $\boldsymbol{E}_{\mathrm{HORZ}}$ actuation. Applied $\boldsymbol{E}_{\mathrm{HORZ}}$ selectively and reversibly accelerates slow inactivation of a voltage-gated potassium channel while leaving activation essentially unchanged. Physical considerations further indicate that $\boldsymbol{E}_{\mathrm{HORZ}}$ arises naturally wherever membrane potential varies spatially, including at action potential wavefronts, suggesting broader physiological relevance. This platform provides experimental access to vector-resolved membrane electric fields and establishes a generalizable strategy for multidirectional electrical control of *soft-matter* biointerfaces.

Maki Komiya, Madoka Sato, Teng Ma, and Hironori Kageyama contributed equally to this work.

## 1. Introduction

Cell membranes are molecularly thin (~4 nm) lipid bilayer shells that enclose living cells and sustain electric fields on the order of $10^7$ V $m^{-1}$ [1–3], similar in magnitude to those routinely employed in solid-state nanoscale devices. This intense field arises because biological membranes maintain voltages of tens to hundreds of millivolts across their nanometer-scale thickness [1–3]. The voltage across the membrane thickness ($V_{VERT}$) gives rise to the corresponding vertical electric field component ($\boldsymbol{E}_{VERT}$). $V_{VERT}$ has long served as the principal electrical parameter governing ion transport and membrane-protein function [1,4–6].

In engineered nanoscale systems, voltages are not restricted to a single direction. Modern solid-state devices routinely exploit orthogonal voltages to modulate charge flow within nanoscale channels, enabling transistor operation and advanced switching functionality. Such voltage configurations generate multidimensional electric fields whose precise control is indispensable for device performance at the nanoscale. Extending analogous multidirectional field control to cell membranes raises a fundamental question: what happens when an electric-field component parallel to the membrane plane is applied to a membrane system that has long been treated as a one-dimensional capacitor?

Implementing such in-plane field control in cell membranes, however, requires overcoming a fundamental geometric constraint. In living cells, the lipid bilayer forms a closed interface separating intracellular and extracellular aqueous compartments, leaving the hydrophobic core entirely enclosed by aqueous phases, with no accessible peripheral region. In standard voltage-clamp configurations, electrodes are placed in the aqueous solutions flanking the membrane [7–9]; while this arrangement enables precise control of $V_{VERT}$, the hydrophobic core remains physically inaccessible from the aqueous phase. Consequently, there is no means to apply a well-defined in-plane voltage within the bilayer core independently of $V_{VERT}$. The planar lipid bilayer (PLB) system addresses this constraint through a fundamentally different geometry. A PLB is formed across an aperture, with its periphery in contact with a hydrophobic organic phase that is continuous with the bilayer's hydrophobic core [10–12]. Electrodes positioned at the aperture edge can therefore access this hydrophobic peripheral region, remaining insulated from the aqueous solutions while coupled to the bilayer interior. This geometry enables independent application of an in-plane voltage ($V_{HORZ}$) alongside $V_{VERT}$, providing directional voltage control otherwise unavailable in conventional cellular electrophysiology [13,14].

To directly investigate the consequences of such an in-plane field component, we established a bioelectronic platform that enables controlled application of $V_{HORZ}$ to a PLB, thereby generating a horizontal electric field ($\boldsymbol{E}_{HORZ}$) within the hydrophobic core. Using this platform, we examined how $\boldsymbol{E}_{HORZ}$ influences lipid bilayer properties and membrane proteins, including ion channels, in a reconstituted, cell-free membrane system. We define $\boldsymbol{E}_{HORZ}$ as the horizontal component of the electric field confined within the dielectric interior of the lipid bilayer.

Beyond engineered control, spatial variations in membrane voltage naturally produce an in-plane field component within the bilayer interior. During action potential propagation in axons, unexcited and excited regions transiently exhibit opposite membrane voltages, creating steep spatial voltage gradients along the membrane plane [15–18]. From this geometry, the vertical electric field component ($\boldsymbol{E}_{VERT}$) must change polarity across the wavefront, implying the emergence of a horizontal electric field component ($\boldsymbol{E}_{HORZ}$) within the lipid bilayer (**Figure 1**). Together, $\boldsymbol{E}_{VERT}$ and $\boldsymbol{E}_{HORZ}$ define the full electric field vector within the bilayer. By enabling independent control of $\boldsymbol{E}_{VERT}$ and $\boldsymbol{E}_{HORZ}$ in a PLB, our platform provides experimental access to vector-resolved membrane electric fields and their consequences for *soft-matter* biointerfaces.

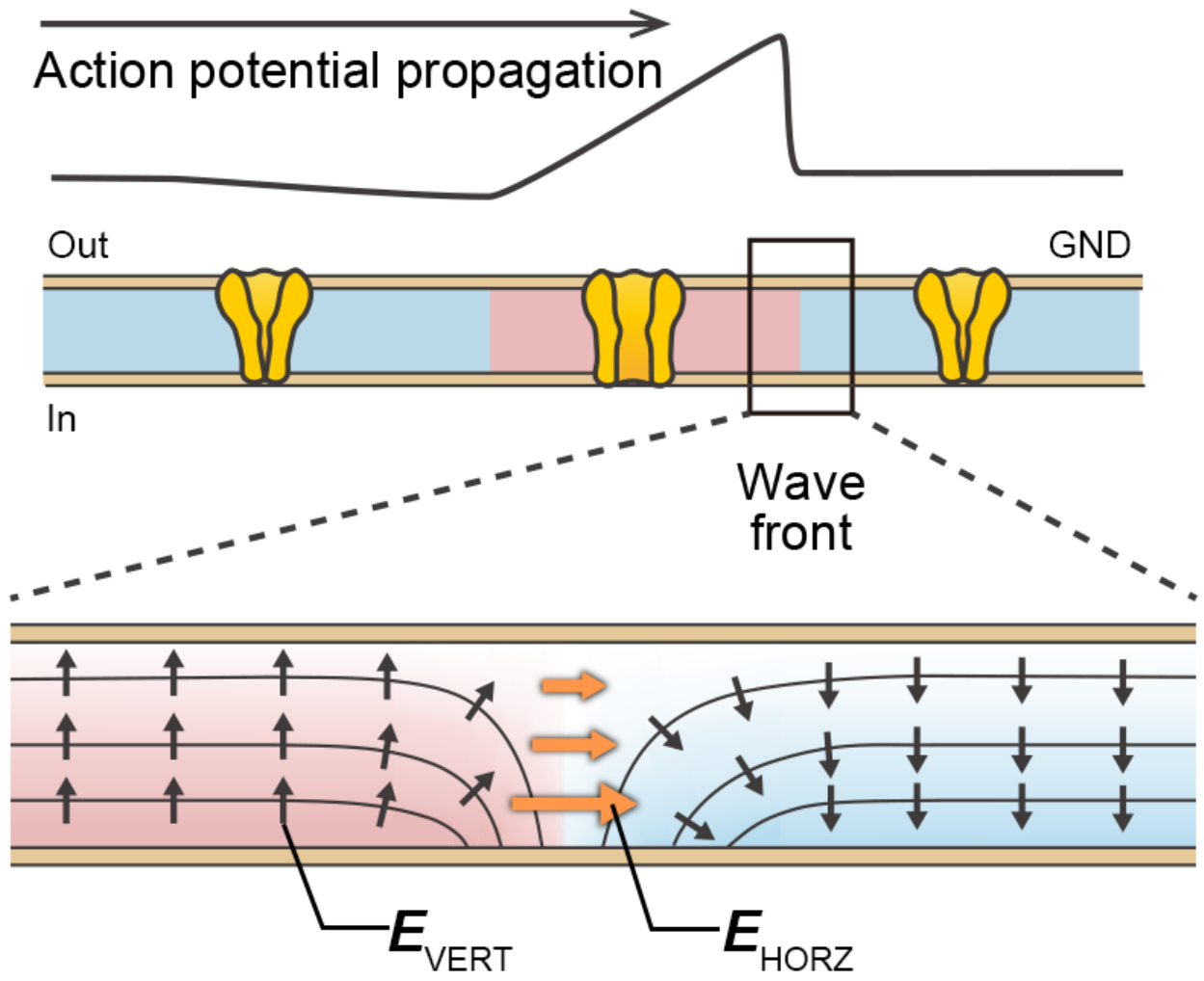

**Figure 1. Origin of the horizontal membrane electric field ($\boldsymbol{E}_{\mathrm{HORZ}}$) during action potential propagation.** As an action potential wavefront travels along the membrane, the vertical electric field ($\boldsymbol{E}_{\mathrm{VERT}}$) in the resting (right, blue) and depolarized (left, red) regions points in opposite directions. At the transition zone between these regions, the spatial gradient of the vertical potential necessarily generates an electric field component parallel to the membrane plane, which corresponds to the horizontal field $\boldsymbol{E}_{\mathrm{HORZ}}$ (see also **Table S1**). This conceptual model highlights that the total electric field within the membrane is three-dimensional.

## 2. Results

### 2.1. Device architecture for the $V_{\mathrm{HORZ}}$ application in a planar lipid bilayer

A central requirement of this bioelectronic platform is the controlled application of horizontal membrane voltage ($V_{\mathrm{HORZ}}$) within a planar lipid bilayer (PLB). To meet this, we developed a $V_{\mathrm{HORZ}}$ system that enables horizontal-field actuation in a PLB (**Figure 2**A) [12,19]. A PLB consists of a nanometer-scale lipid bilayer surrounded by a micrometer-scale organic solvent annulus known as the torus (Figure 2B and **Figure 3**A) [10]. Our device design strategically exploits this geometry by patterning a pair of micrometer-scale electrodes at opposite edges of the aperture (Figure 2C) such that, upon PLB formation, the electrodes do not contact the aqueous phases but instead lie within the torus (Figure 2B). Because the torus is continuous with the bilayer's hydrophobic core, it serves as a critical interface that couples the applied $V_{\mathrm{HORZ}}$ to the membrane interior. This configuration is designed to concentrate the $\boldsymbol{E}_{\mathrm{HORZ}}$ component within the membrane core and to act across a lipid bilayer of ~100 μm in diameter and ~4 nm thickness, thereby bridging the large disparity between the *hard*-material device scale and the nanometer-scale *soft-matter* membrane interior (Figure 2B).

### 2.2. Fabrication and electrical validation of the $V_{\mathrm{HORZ}}$ electrode chip

To realize $\boldsymbol{E}_{\mathrm{HORZ}}$ actuation in a PLB, the $V_{\mathrm{HORZ}}$ electrode chip must combine a geometry that couples the applied voltage to the membrane interior and teraohm-level insulation that suppresses ionic leakage through the aqueous phases. Figure 2D outlines the fabrication process of the $V_{\mathrm{HORZ}}$ electrode chip. After forming a microaperture (70-180 μm in diameter) in a Teflon sheet, two Ti electrodes were deposited in parallel around the aperture by electron-beam evaporation. The distance between the two electrodes was 49.8 ± 0.9 μm ($n$ = 53, mean ± s.e.m.). The Ti-facing surface of the Teflon sheet was coated with $SiO_2$. The Ti and $SiO_2$ layers extended to approximately the midpoint of the aperture wall, corresponding to the narrowest region in the aperture (Figure 2E and **Figure S4**A). This geometry is designed to promote field coupling while minimizing direct current leakage into the aqueous phases. The exposed Ti surfaces were then covered with Pt to prevent oxidation.

The insulation properties of the $V_{\mathrm{HORZ}}$ electrode chip were assessed by measuring the leakage current when the chip, immersed in electrolyte, was connected in series with a $V_{\mathrm{HORZ}}$ source and an ammeter (Figure 2F). In the unmodified chip, the leakage current through the electrolyte was ~4 pA at a DC $V_{\mathrm{HORZ}}$ of 4 V, corresponding to a resistance of ~1 TΩ (Figure 2F, blue). After surface modification with the fluorosilane PFDS ((1H,1H,2H,2H-perfluorooctyl)-dimethylchlorosilane; see Supplementary Information (SI)), the resistance increased to ~6 TΩ (Figure 2F, green). This teraohm-level insulation is essential because it confines the applied $V_{\mathrm{HORZ}}$ to the hydrophobic core of the bilayer, generating a pure $\boldsymbol{E}_{\mathrm{HORZ}}$ without undesirable ionic leakage through the surrounding electrolyte. These measurements establish teraohm-level electrical isolation of the $V_{\mathrm{HORZ}}$ path under aqueous conditions, a prerequisite for directional field delivery to the membrane interior.

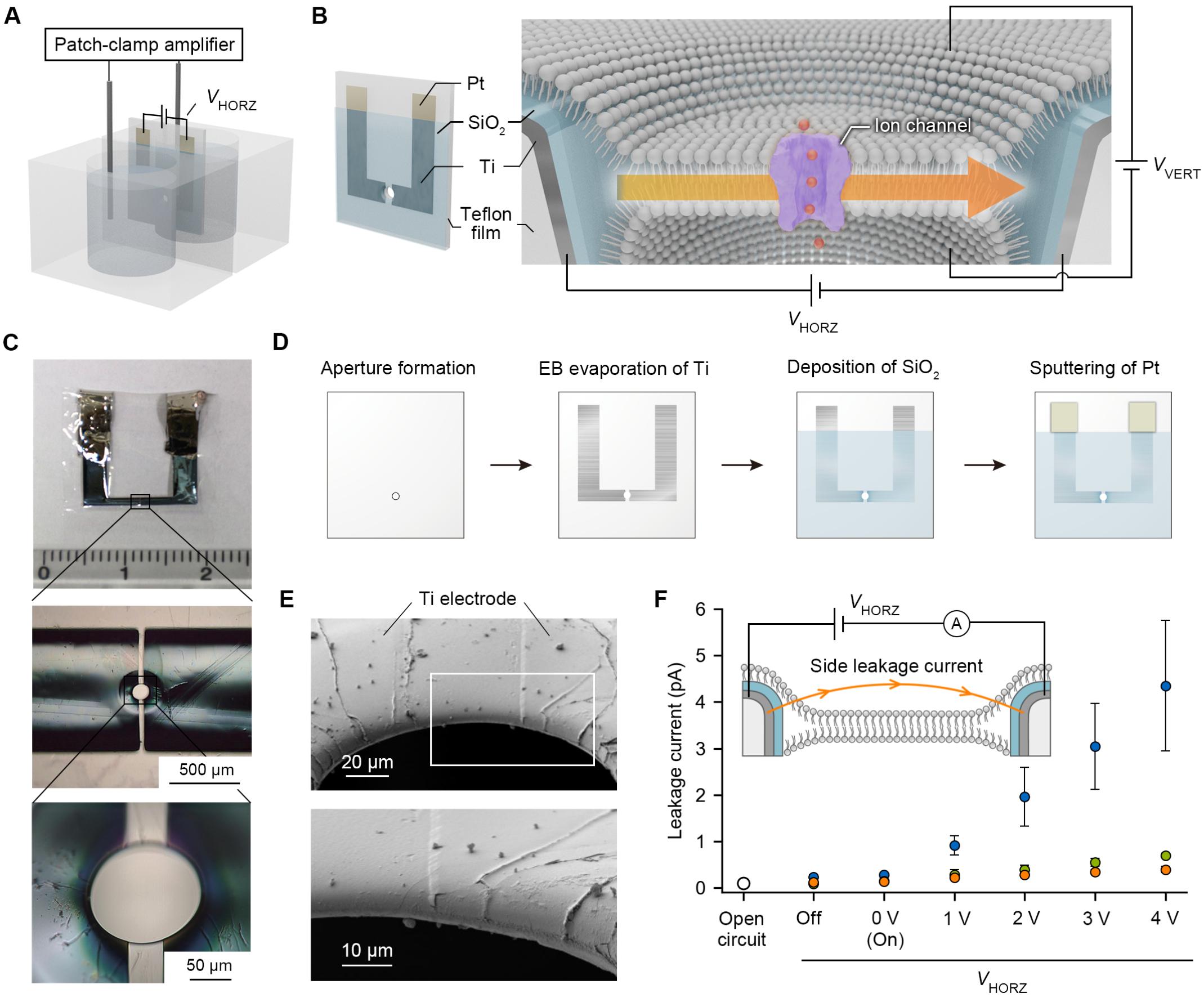

**Figure 2. Design, fabrication, and electrical validation of the $V_{\mathrm{HORZ}}$ electrode chip.** **(A)** Schematic of the experimental setup, enabling the simultaneous and orthogonal application of transmembrane voltage ($V_{\mathrm{VERT}}$) and in-plane horizontal voltage ($V_{\mathrm{HORZ}}$). **(B)** $V_{\mathrm{HORZ}}$ electrode chip and cross-sectional illustration of the ion-channel-incorporated planar lipid bilayer (PLB) with orthogonal electrode sets (not to scale). A pair of electrodes is embedded within the lipid torus at the borders of the PLB, ensuring electrical isolation from the aqueous electrolyte and positioning them to generate a horizontal electric field ($\boldsymbol{E}_{\mathrm{HORZ}}$) within the membrane's dielectric interior. **(C)** Photographs of the $V_{\mathrm{HORZ}}$ electrode chip. **(D)** Overview of the chip fabrication process. **(E)** Scanning electron microscopy (SEM) images of the fabricated chip around the aperture and the deposited electrode structure. **(F)** Electrical validation of the chip's insulation properties. Leakage current across the $V_{\mathrm{HORZ}}$ electrodes was measured in electrolyte solution (mean ± s.e.m., $n = 3$). Surface modification with PFDS (green) substantially reduces the leakage current compared with the unmodified chip (blue) ($n = 3$), yielding a resistance of approximately 6 TΩ. This teraohm-level insulation confines the applied $V_{\mathrm{HORZ}}$ to the hydrophobic membrane core, generating a predominantly pure $\boldsymbol{E}_{\mathrm{HORZ}}$ with minimal ionic leakage through the surrounding solution. The resistance remained high after bilayer formation (orange). The measurement noise level is shown in white ($n = 12$).

### 2.3. Lipid bilayer formation in a $V_{\mathrm{HORZ}}$ electrode chip

To use the $V_{\mathrm{HORZ}}$ electrode chip as a bioelectronic platform, it must support stable PLB formation and maintain electrical isolation between $V_{\mathrm{HORZ}}$ actuation and conventional $V_{\mathrm{VERT}}$ stimulation. A PLB was formed from a 1,2-dioleoyl-sn-glycero-3-phosphocholine (DOPC) solution (DOPC:cholesterol = 4:1 by weight) using the folding method, with *n*-hexadecane as a pre-coating solvent [10,19,20]. To apply $V_{\mathrm{VERT}}$, a pair of conventional electrodes was placed in the electrolyte (Figure 2A). Visual inspection revealed a clear boundary between the torus and lipid bilayer, with the bilayer occupying 60-90% of the aperture area (**Figure 3**A (i), right). The membrane resistance of the PLBs in response to $V_{\mathrm{VERT}}$ exceeded 200 GΩ (Figure S4C), indicating that the Ti electrodes and $SiO_2$ layer around the microaperture did not compromise bilayer stability.

In the presence of a lipid bilayer, the $V_{\mathrm{HORZ}}$ electrodes are insulated from the aqueous phases because they reside within the torus, which is continuous with the hydrophobic core of the bilayer. To evaluate this insulation, we measured side-leakage currents (Figure 2F, inset), defined as currents passing through the organic phase and lipid monolayer and into the electrolyte during steady $V_{\mathrm{HORZ}}$ application. These leakage currents (Figure 2F, orange) were negligible compared with those observed in the absence of a membrane, indicating that the teraohm-level insulation effectively confines the applied $V_{\mathrm{HORZ}}$ to the bilayer interior. However, during prolonged $V_{\mathrm{HORZ}}$ application, the contact resistance of the electrodes, defined as the resistance of the Pt-coated Ti region, increased owing to oxidation of the underlying Ti (Figure S4B). In subsequent experiments, the duration of the $V_{\mathrm{HORZ}}$ application was therefore limited to ~30 min to keep the contact resistance below the kilo-ohm range. We also examined whether $V_{\mathrm{HORZ}}$ perturbs vertical leakage currents measured at $V_{\mathrm{VERT}}$ = +100 mV (Figure S4C). The vertical leakage currents observed in the presence of $V_{\mathrm{HORZ}}$ (4 V DC) were negligible and indistinguishable from those recorded without $V_{\mathrm{HORZ}}$. Together, these results demonstrate that $V_{\mathrm{HORZ}}$ and $V_{\mathrm{VERT}}$ are electrically isolated under our experimental conditions, enabling orthogonal control of electric field orientation within the membrane interior.

### 2.4. $\boldsymbol{E}_{\mathrm{HORZ}}$ actuation preserves membrane thickness

Next, we evaluated the specific capacitance ($C_{\mathrm{sp}}$, μF cm$^{-2}$) of the lipid bilayer **(Figure 3A** and **Figure S1)** [21] by applying a continuous ramp $V_{\mathrm{VERT}}$ [10,12,22]. The bilayer area ($A_{\mathrm{m}}$) was determined through image analysis, and $C_{\mathrm{sp}}$ was calculated by dividing the capacitance of the lipid bilayer ($C_{\mathrm{m}}$) by its area ($C_{\mathrm{sp}} = C_{\mathrm{m}} / A_{\mathrm{m}} = \varepsilon_0\varepsilon_{\mathrm{m}} / d$, where $\varepsilon_0$, $\varepsilon_{\mathrm{m}}$, and $d$ are the vacuum permittivity, relative permittivity of the lipid bilayer, and membrane thickness, respectively) (see SI, Equation 1). Without $V_{\mathrm{HORZ}}$, $C_{\mathrm{sp}}$ was 0.64 ± 0.04 μF cm$^{-2}$ at a $V_{\mathrm{VERT}}$ of +100 mV (mean ± s.e.m., $n$ = 15) (Figure 3A), consistent with reported $C_{\mathrm{sp}}$ values for bilayers with similar lipid composition [23–25]. Upon application of $V_{\mathrm{HORZ}}$, $C_{\mathrm{sp}}$ remained unchanged (0.64 ± 0.04 μF cm$^{-2}$), indicating that the $\boldsymbol{E}_{\mathrm{HORZ}}$ actuation does not measurably affect membrane thickness under our experimental conditions.

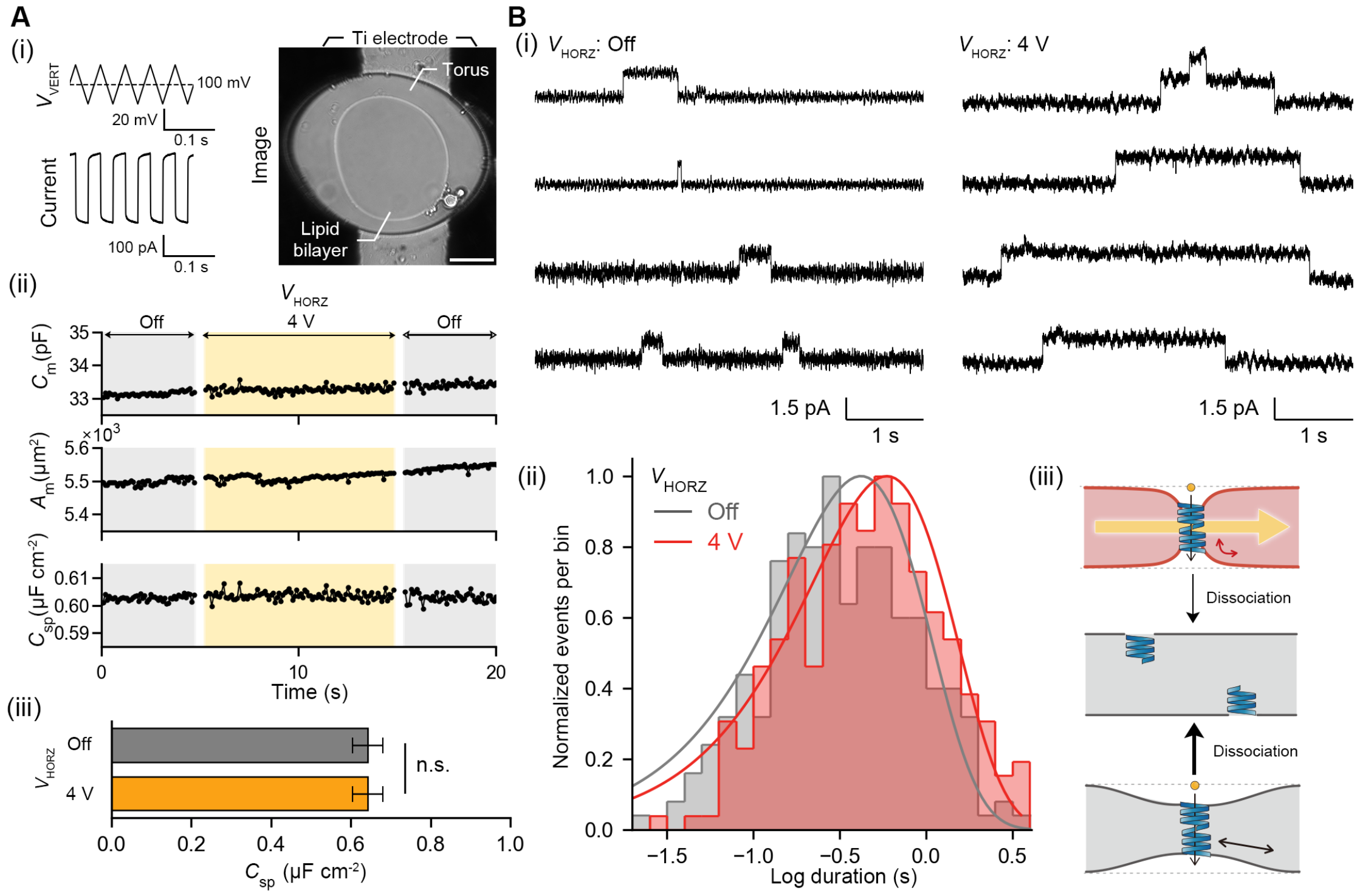

**Figure 3. $V_{\mathrm{HORZ}}$ application alters the mechanical properties of the lipid bilayer without changing its thickness. (A)** Membrane thickness remains constant under $V_{\mathrm{HORZ}}$ application, as determined by membrane specific capacitance ($C_{\mathrm{sp}}$) measurements. $C_{\mathrm{sp}}$ with and without $V_{\mathrm{HORZ}}$ at $V_{\mathrm{VERT}}$ = +100 mV. (i) Principles of the $C_{\mathrm{sp}}$ measurements. $C_{\mathrm{sp}}$ was evaluated by simultaneous measurements of the lipid bilayer capacitance ($C_{\mathrm{m}}$) and the bilayer area ($A_{\mathrm{m}}$), providing a direct probe of membrane thickness. $C_{\mathrm{m}}$ was evaluated by applying a $V_{\mathrm{VERT}}$ ramp protocol at +100 mV, with stray capacitance subtracted (see Figure S1). $A_{\mathrm{m}}$ was evaluated by microscopy. $C_{\mathrm{sp}} = C_{\mathrm{m}} / A_{\mathrm{m}} = \varepsilon_0 \varepsilon_{\mathrm{m}} / d$, where $\varepsilon_0$, $\varepsilon_{\mathrm{m}}$, and $d$ are the permittivity of vacuum, relative permittivity of the lipid bilayer, and membrane thickness, respectively. Scale bar: 25 μm. (ii) Representative time course of $C_{\mathrm{m}}$, $A_{\mathrm{m}}$, and $C_{\mathrm{sp}}$, showing the stability of $C_{\mathrm{sp}}$ before, during, and after the application of $V_{\mathrm{HORZ}}$. (iii) Summary of $C_{\mathrm{sp}}$ values. No statistically significant change was observed upon 4 V of $V_{\mathrm{HORZ}}$ application (yellow; $p > 0.05$, two-sided paired $t$-test, $n = 15$). **(B)** $V_{\mathrm{HORZ}}$ modulates membrane mechanical properties, as probed by the lifetime of gramicidin A (gA) channels. (i) Representative gA single-channel current traces recorded at $V_{\mathrm{VERT}}$ = +100 mV in the absence (Off) and presence (4 V) of $V_{\mathrm{HORZ}}$. (ii) Lifetime distributions for gA channels with (4 V: red) and without (Off: gray) $V_{\mathrm{HORZ}}$. The lifetime distributions are plotted as log-binned histogram [87] and fitted by single exponential functions, $N(t) = a \exp[\ln(t) - \ln(\tau) - \exp\{\ln(t) - \ln(\tau)\}]$, where $N(t)$ is the number of channels with lifetime $t$, $\tau$ is the average single-channel lifetime, and $a$ is the scaling factor [87]. $V_{\mathrm{HORZ}}$ significantly prolonged $\tau$ from 0.40 ± 0.04 s to 0.62 ± 0.05 s (mean ± s.e.m., $n = 4$) ($p < 0.05$, independent two-sample $t$-test), indicating a stabilization of the gA dimer. (iii) Mechanistic interpretation. The gA dimer is shorter than the membrane, inducing a local deformation that creates mechanical stress and promotes dimer dissociation. The $V_{\mathrm{HORZ}}$ application alters the local mechanical state of the bilayer in a way that reduces the effective deformation-induced stress experienced by the channel. This reduced mechanical penalty stabilizes the dimer, yielding a longer lifetime. The yellow arrow indicates the $\boldsymbol{E}_{\mathrm{HORZ}}$.

## 2.5. $E_{\mathrm{HORZ}}$ actuation modulates membrane mechanical properties

We investigated how $V_{\mathrm{HORZ}}$ affects the mechanical properties of lipid bilayers using the gramicidin A (gA) channel [26], a peptide dimer widely used as a sensitive molecular force probe [27,28]. Formation of a conducting gA dimer requires the surrounding lipid bilayer to deform locally to match the shorter hydrophobic span of the protein channel [26,29,30]. This deformation incurs an energetic cost ($\Delta G_{def}$) determined by the hydrophobic mismatch and the elastic moduli of the membrane (see SI); consequently, the dimer open-state lifetime provides a direct readout of the bilayer energetic landscape [27,31–34].

In the absence of $V_{HORZ}$, the average open-state lifetime ($\tau$) was 0.40 ± 0.04 s at $V_{VERT}$ = +100 mV, consistent with reported values under similar conditions [35,36]. Upon application of $V_{HORZ}$ (4 V DC), $\tau$ significantly increased to 0.62 ± 0.05 s ($p < 0.05$), indicating a stabilization of the dimer state (Figure 3B). Crucially, the specific capacitance remained unchanged under $V_{HORZ}$ (Figure 3A), indicating that membrane thickness is not measurably altered and allowing us to exclude reduced hydrophobic mismatch as the primary cause of the stabilization.

Instead, interpreting these results within the continuum elastic framework established by Andersen and colleagues [31,34,37–43], we attribute the lifetime prolongation to a field-induced modulation of the bilayer's intrinsic curvature ($c_0$) and/or bending rigidity ($k_c$) (see SI) [28]. Because $\boldsymbol{E}_{HORZ}$ acts orthogonal to lipid dipoles, it can alter headgroup orientation and lateral packing interactions [44]. Such modifications can shift the membrane's spontaneous curvature or stiffness in a manner that reduces the deformation energy penalty associated with gA dimer formation. This finding suggests that $\boldsymbol{E}_{HORZ}$ does not merely act as an external force field but actively tunes the mechanical properties of the lipid environment, thereby modulating the function of mechanosensitive membrane proteins.

### 2.6. Optical spatial mapping reveals distributed $V_{HORZ}$ effects across the lipid bilayer

We next employed ratiometric fluorescence imaging with the voltage-sensitive dye di-4-ANEPPS (**Figure 4**A and 4B) [45–48] to quantify the spatial extent of $\boldsymbol{E}_{HORZ}$ influence. A critical question for this device architecture is whether the horizontal-field actuation delivered at the torus-embedded electrodes affects the macroscopic lipid bilayer uniformly or exhibits measurable decay with distance from the rim.

We first mapped the fluorescence excitation ratio, $R_{ex}$ ($F_{435}$ / $F_{500}$), across the lipid bilayer while clamping at a series of $V_{VERT}$. In the absence of $V_{HORZ}$, $R_{ex}$ was spatially homogeneous across the entire membrane and increased linearly with $V_{VERT}$ (**Figure S2**), allowing determination of a $R_{ex}$-$V_{VERT}$ slope for each pixel as the local baseline sensitivity to $V_{VERT}$ [45,47,49]. We then applied a constant $V_{HORZ}$ and calculated, for each pixel, the ratio of the $R_{ex}$-$V_{VERT}$ slope with $V_{HORZ}$ to its baseline value.

The spatial analysis yielded two insights. First, the modulation of dye response was spatially distributed across the membrane. The slope ratio was flat on average along $x$ from the positive $V_{HORZ}$ electrode side to the opposing side of the bilayer (Figure 4C and 4E), consistent with effective confinement of $V_{HORZ}$ to the membrane interior and spatially distributed $V_{HORZ}$ coupling with no systematic distance-dependent decay, despite trial-to-trial variability. Second, contrary to the expectation of a simple superposition, we observed a systematic enhancement in the dye's voltage sensitivity under $V_{HORZ}$ application. The mean slope ratio was consistently greater than unity (1.05 ± 0.01, $n$ = 7, $p < 0.001$) across the bilayer under the application of $V_{HORZ}$, whereas no detectable difference was observed in its absence (0.99 ± 0.01, $n$ = 7; Figure 4D and **Figure S5**). Since membrane thickness remains constant (Figure 3A), this steepening of the $R_{ex}$-$V_{VERT}$ relationship cannot be attributed to an increase in $\boldsymbol{E}_{VERT}$ magnitude. Rather, this phenomenon is best explained by considering the three-dimensional nature of the electric field interaction. Di-4-ANEPPS is an electrochromic dye with a molecular dipole moment that is tilted relative to the membrane normal [50–52], and applying $V_{HORZ}$ introduces a horizontal field component such that the total electric field vector ($\boldsymbol{E}_{TOTAL} = \boldsymbol{E}_{VERT} + \boldsymbol{E}_{HORZ}$) becomes oblique within the membrane. The observed sensitivity enhancement suggests that this resultant field vector aligns more favorably with the dye's dipole moment than $\boldsymbol{E}_{VERT}$ alone, thereby increasing the efficiency of the Stark shift (Figure 4F). Thus, our optical mapping not only establishes spatially distributed $V_{HORZ}$ coupling across the PLBs but also reveals its capacity to modulate the electro-optical properties of membrane-bound molecules through vectorial field summation.

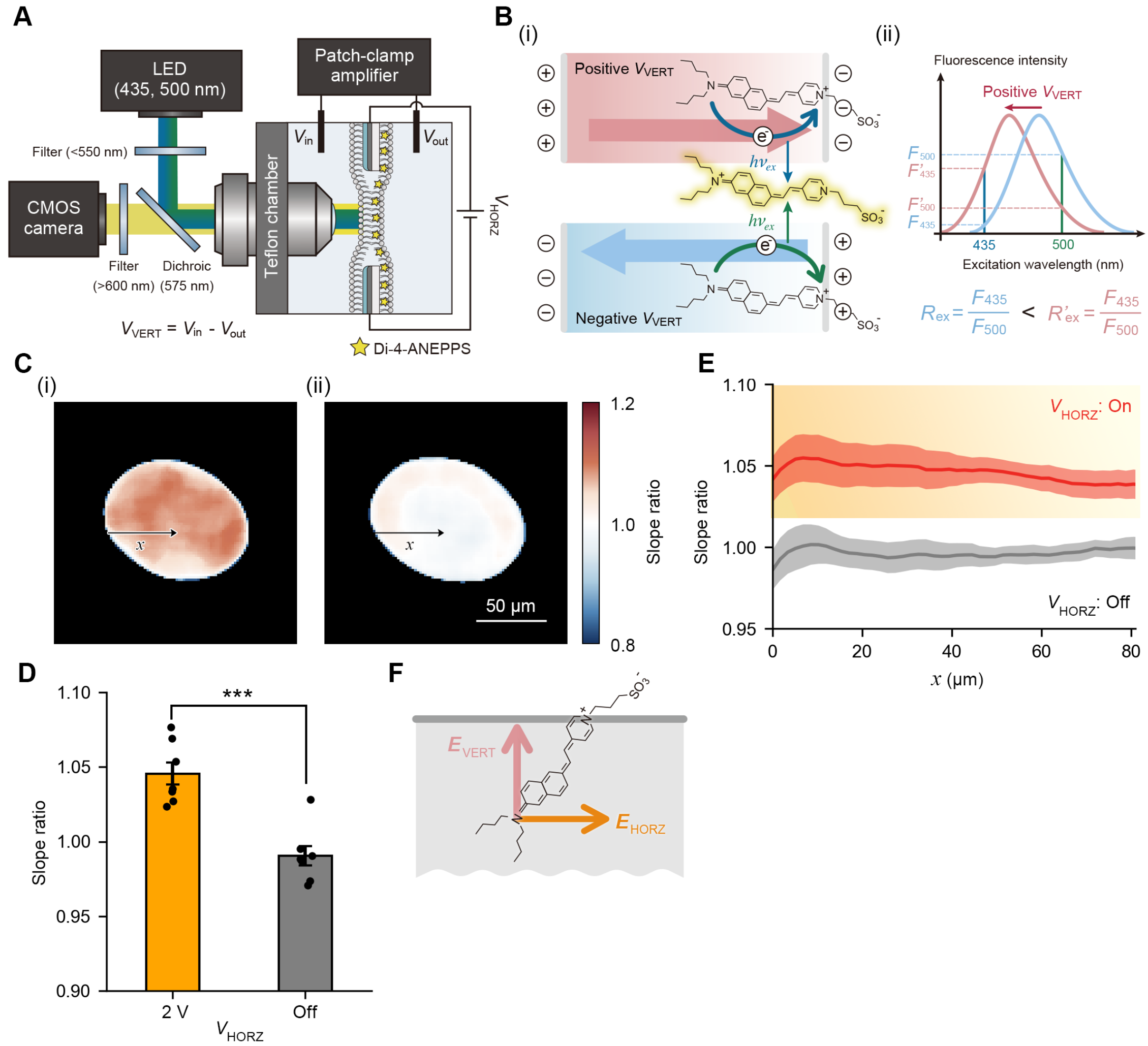

**Figure 4. Spatial distribution of $V_{HORZ}$ effects across the lipid bilayer.** (**A**) Schematic of the ratiometric fluorescence imaging setup for a PLB. (**B**) Chemical structure and spectral properties of the voltage-sensitive dye, di-4-ANEPPS. $R_{ex}$ denotes the ratio of fluorescence intensities excited at 435 and 500 nm. (**C**) Maps of the $R_{ex}$-$V_{VERT}$ slope ratio in the presence (i) and absence (ii) of $V_{HORZ}$, obtained from the same PLB. Ratios greater than unity indicate a steeper $R_{ex}$-$V_{VERT}$ relation under $V_{HORZ}$. In the control condition (ii), two successive slope measurements were performed in the absence of $V_{HORZ}$. (**D**) Quantification of the slope ratio across PLBs. Application of $V_{HORZ}$ increased the mean slope ratio (mean ± s.e.m., $n = 7$ PLBs; $p < 0.001$, two-sided paired $t$-test). (**E**) Spatial distribution of the slope ratio. The $R_{ex}$-$V_{VERT}$ slope ratio is plotted as a function of distance $x$, defined along an axis running from the side of the lipid bilayer closer to the positive $V_{HORZ}$ electrode toward the opposite (see panel C for orientation). In panels **C-E**, $V_{HORZ} = 2$ V. (**F**) Schematic illustrating that the dipole moment of di-4-ANEPPS has components along both $\boldsymbol{E}_{VERT}$ and $\boldsymbol{E}_{HORZ}$, providing a possible basis for the observed modulation of the $R_{ex}$-$V_{VERT}$ relation.

## 2.7. $E_{\mathrm{HORZ}}$ actuation accelerates the inactivation of a voltage-gated potassium channel

As a functional demonstration of this vector-resolved actuation platform, we examined whether $\boldsymbol{E}_{\mathrm{HORZ}}$ modulates the gating behavior of a voltage-gated ion channel, a protein machine essential for physiological excitability. To address this, we reconstituted the archetypal voltage-gated potassium channel, KvAP, into the PLB [53–56] and analyzed its gating behavior under the application of $V_{\mathrm{HORZ}}$ (see Methods). Purified KvAP channels were reconstituted into a PLB of diphytanoylphosphatidylcholine (DPhPC) [57]. Macroscopic currents provided a direct readout of the gating conformational changes in the channels under a series of $V_{\mathrm{VERT}}$ protocols (see Methods and Notes in SI). KvAP undergoes depolarization-activated movement of the voltage sensor domains (VSDs), followed by opening of the activation gate (**Figure 5**A and 5B). Meanwhile, the inactivation gate closes slowly when $V_{\mathrm{VERT}}$ is prolonged (C-type inactivation) [58,59], preventing the reopening of the channel (Figure 5A and 5D; see SI for the gating characterization). These characteristic signatures reproduce previously reported KvAP gating behaviors [55,60,61].

We first examined whether the $V_{\mathrm{HORZ}}$ application affects $V_{\mathrm{VERT}}$-dependent activation, the primary voltage-sensing process. Even under a strong, steady $V_{\mathrm{HORZ}}$ (5 V), activation was essentially unchanged (Figure 5B). The conductance-voltage ($G$-$V_{\mathrm{VERT}}$) relationship (Figure 5C), fitted with a Boltzmann function (Methods), showed no detectable shift, indicating that the energy required to open the activation gate [62] was unaffected. Likewise, the activation kinetics, represented as $\tau^{\mathrm{Act}}$, were negligibly altered across $V_{\mathrm{VERT}}$ values (Figure 5E). These observations indicate that $\boldsymbol{E}_{\mathrm{HORZ}}$ does not perturb the canonical movement of the VSDs, which detect $\boldsymbol{E}_{\mathrm{VERT}}$ and initiate channel opening (Figure 5A) [5,63].

In sharp contrast, the $V_{\mathrm{HORZ}}$ application had a strong and reversible effect on the slow C-type inactivation. With long depolarizing pulses (Figure 5D), $V_{\mathrm{HORZ}}$ markedly accelerated the decay of macroscopic currents. This acceleration was observed consistently across a range of $V_{\mathrm{VERT}}$, resulting in significantly reduced inactivation time constants $\tau^{\mathrm{Inact}}$ (Figure 5E) [64,65]. Importantly, this acceleration was fully reversible upon $V_{\mathrm{HORZ}}$ removal (Figure 5D (ii)), indicating a specific modulatory effect rather than an irreversible denaturation or damage to the channel. $V_{\mathrm{HORZ}}$ also produced negligible changes in steady-state inactivation (Figure 5F and **Figure S6**), showing that the inactivated state itself is not stabilized or destabilized; rather, only the kinetics of entering the inactivated state are accelerated.

This differential modulation—sparing VSD-driven activation while potently accelerating PD-mediated C-type inactivation [66–69]—provides insight into the underlying mechanism. It suggests that $\boldsymbol{E}_{\mathrm{HORZ}}$ does not act on the VSDs like $\boldsymbol{E}_{\mathrm{VERT}}$ does. Instead, it selectively influences the conformational landscape of the pore domain, either by directly acting on the protein or, consistent with our gramicidin A findings, by altering the mechanical forces exerted by the lipid bilayer on the pore. This represents an orthogonal axis of control for ion channel function.

The ability to accelerate inactivation without altering the $V_{\mathrm{VERT}}$ threshold for activation has meaningful physiological implications. In neurons, the rate of potassium channel inactivation is a major determinant of action-potential duration and maximal firing frequency [1]. Our findings raise the possibility that intrinsic $\boldsymbol{E}_{\mathrm{HORZ}}$ within biological membranes could contribute to fine-tuning neuronal excitability and information processing, complementing the canonical role of $\boldsymbol{E}_{\mathrm{VERT}}$.

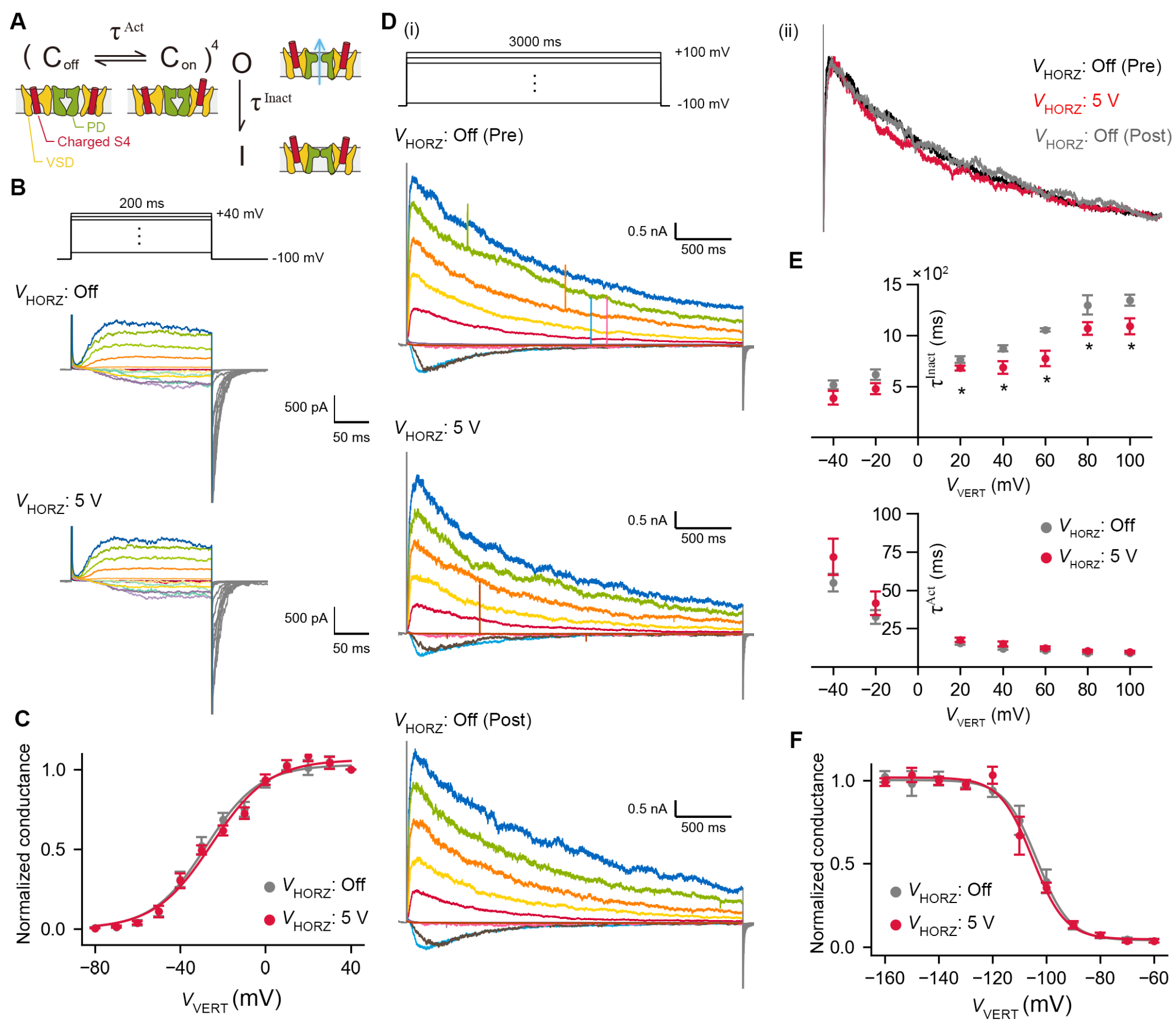

**Figure 5. $V_{\mathrm{HORZ}}$ selectively and reversibly accelerates the inactivation kinetics of the KvAP channel.** (**A**) A schematic representation of voltage-gated potassium (Kv) channel gating. At negative resting membrane potentials, the charged S4 helix in the voltage-sensor domain (VSD) resides in the downward position. Depolarization drives S4 upward, opening the activation gate in the pore domain (PD). Prolonged depolarization leads to slow closure of the inactivation gate in the PD (C-type inactivation). (**B**) $V_{\mathrm{HORZ}}$ has negligible influence on activation gating. Representative current traces recorded during short (200 ms) depolarizing pulses (–90 mV to +40 mV; inset) from a holding $V_{\mathrm{VERT}}$ of –100 mV show that the activation phase is unchanged by a sustained $V_{\mathrm{HORZ}}$ of 5 V. (**C**) The $V_{\mathrm{VERT}}$ dependence of activation is unaffected by $V_{\mathrm{HORZ}}$. Conductance-voltage ($G$-$V_{\mathrm{VERT}}$) relationships obtained in the absence (control) and presence of 5 V $V_{\mathrm{HORZ}}$ nearly overlap. Curves were fitted with a Boltzmann function. Parameters for control were: $V_{\mathrm{half}}^{\mathrm{Act}}$ = –27.9 mV, $k^{\mathrm{Act}}$ = 12.6 mV. Parameters for $V_{\mathrm{HORZ}}$ were: $V_{\mathrm{half}}^{\mathrm{Act}}$ = –25.2 mV, $k^{\mathrm{Act}}$ = 13.4 mV. (**D**) $V_{\mathrm{HORZ}}$ selectively and reversibly accelerates slow inactivation. (i) Representative current traces from a long (3 s) depolarizations show markedly faster decay during $V_{\mathrm{HORZ}}$ application, which is reversed after $V_{\mathrm{HORZ}}$ removal. (ii) Superimposed traces at +100 mV $V_{\mathrm{VERT}}$ highlight the reversible acceleration of inactivation by $V_{\mathrm{HORZ}}$ (red) compared with before (black) and after (gray). (**E**) Acceleration of inactivation by $V_{\mathrm{HORZ}}$ is evident over a wide $V_{\mathrm{VERT}}$ range. Time constants for activation ($\tau^{\mathrm{Act}}$) and inactivation ($\tau^{\mathrm{Inact}}$) are plotted versus $V_{\mathrm{VERT}}$. $\tau^{\mathrm{Act}}$ is minimally altered by $V_{\mathrm{HORZ}}$, whereas $\tau^{\mathrm{Inact}}$ is significantly reduced. *$p < 0.05$ (paired $t$-test, $n = 5$ PLBs). Time constants were obtained by fitting current traces with the Hodgkin-Huxley equation (see Methods). **(F)** Steady-state inactivation is insensitive to $V_{\mathrm{HORZ}}$. Steady-state inactivation curves, constructed by plotting peak current at $V_{\mathrm{VERT}}$ = +100 mV against the holding potential (see **Figure S6**), show no significant shift. $V_{\mathrm{half}}^{\mathrm{Inact}}$ was –102.7 mV and $k^{\mathrm{Inact}}$ was –6.78 mV for the control, and $V_{\mathrm{half}}^{\mathrm{Inact}}$ was –104.2 mV and $k^{\mathrm{Inact}}$ was –6.87 mV for a $V_{\mathrm{HORZ}}$ of 5 V.

## 3. Discussion

In this study, we established an experimental platform to investigate the fundamental consequences of $\boldsymbol{E}_{\mathrm{HORZ}}$ using a well-defined, reconstituted model membrane. This bottom-up strategy allowed us to dissect the core physical principles governing $\boldsymbol{E}_{\mathrm{HORZ}}$-membrane interactions, free from the inherent complexities of a living cell membrane. To achieve this, we developed a device that exploits the distinctive geometry of the PLB system, in which a micrometer-scale lipid torus is continuous with the nanometer-scale lipid bilayer. By embedding micrometer-scale electrodes within the torus, we created an interface designed to couple and confine the applied electric field from the *hard* micro-electrodes to the *soft* nanometer-scale membrane interior (Figure 2), enabling vector-resolved field actuation under aqueous conditions. This technological foundation enabled a cascade of discoveries. First, by using a voltage-sensitive dye, we visualized how the influence of $V_{\mathrm{HORZ}}$ spreads evenly across the lipid bilayer (Figure 4). These optical measurements further suggest that the dye reports not $\boldsymbol{E}_{\mathrm{VERT}}$ alone but the resultant three-dimensional field vector within the membrane. This insight has broad implications for the interpretation of voltage-sensitive fluorescence signals. Second, using gramicidin A channels as molecular force probes, we found that $\boldsymbol{E}_{\mathrm{HORZ}}$ modifies the membrane mechanical properties while leaving its thickness unchanged (Figure 3). Although the precise mechanism remains to be determined, these results demonstrate that $\boldsymbol{E}_{\mathrm{HORZ}}$ can alter the physical context of the membrane in which embedded proteins operate. Together, these findings set the stage for our central observation: selective and reversible modulation of a voltage-gated ion channel by $\boldsymbol{E}_{\mathrm{HORZ}}$.

The differential modulation of KvAP gating provides strong evidence for the physiological relevance of $\boldsymbol{E}_{\mathrm{HORZ}}$. While activation driven by $V_{\mathrm{VERT}}$ was largely unaffected, slow C-type inactivation was markedly and reversibly accelerated (Figure 5). This functional decoupling is remarkable. Recent structural studies show that C-type inactivation involves conformational rearrangements of the selectivity filter [67–69]. Although the detailed coupling pathway remains to be established, our findings suggest that $\boldsymbol{E}_{\mathrm{HORZ}}$ acts preferentially on the PD, either by directly perturbing its conformation or by modifying the membrane mechanical forces. Such selective modulation has profound physiological implications. In neurons, the rate and extent of $K^+$ channel inactivation are critical determinants of action potential duration and the capacity for sustained high-frequency firing [1]. $\boldsymbol{E}_{\mathrm{HORZ}}$ generated at the action potential wavefront could therefore contribute to local, activity-dependent fine-tuning of excitability, for instance by modulating spike-frequency adaptation [70], without altering the $V_{\mathrm{VERT}}$ threshold for activation. Conceptually, this represents an orthogonal mode of ion-channel control, arising from the fact that $\boldsymbol{E}_{\mathrm{HORZ}}$ is physically orthogonal to $\boldsymbol{E}_{\mathrm{VERT}}$. At the same time, this mechanism is conceptually analogous to known effects of extracellular current flow on action potential propagation [8].

Our results support a broader re-evaluation of $\boldsymbol{E}_{\mathrm{HORZ}}$ not as an artificial stimulus, but as a plausible and widely occurring component of membrane bioelectricity whenever membrane potential ($V_{\mathrm{VERT}}$) varies along the membrane plane. $\boldsymbol{E}_{\mathrm{HORZ}}$ is not confined to the wavefront of action potentials in axons [15]. $\boldsymbol{E}_{\mathrm{HORZ}}$ is inherent to any biological process involving spatial gradients in membrane potential, such as during synaptic integration in dendrites [71,72] or across the tight junctions of epithelial tissues [73]. Quantitatively, the $\boldsymbol{E}_{\mathrm{HORZ}}$ magnitudes applied here ($10^3$–$10^4$ V m$^{-1}$) align closely with estimates for native biological settings. For instance, $\boldsymbol{E}_{\mathrm{HORZ}}$ values on the order of ~$10^3$ V m$^{-1}$ arise at the action potential wavefront in unmyelinated axons [15], axon initial segment [16,17], and at the node of Ranvier in myelinated axons [18] (see SI, **Table S1**, Equation 6-8, **Figure S8** and **Figure S9**). Steady-state fields of ~$10^3$ V m$^{-1}$ also occur across epithelial tight junctions due to transepithelial potentials [73–77]. This quantitative correspondence directly links our *in vitro* platform to *in vivo* physiology, indicating that the effects we observe occur within field strengths naturally present in cells and tissues.

Historically, the action potential has been viewed primarily as an electrical event, yet it is intrinsically coupled to mechanical changes. Tasaki's classic experiments demonstrated that axonal excitation involves reversible swelling and shrinkage, but the molecular actuator driving these changes has remained undefined [78,79]. El Hady and Machta supported Tasaki's finding with sophisticated methods and proposed an electrostriction model [80]. While the thermodynamic and macroscopic reality of these mechanical waves is undeniable, their specific molecular origin has remained elusive, often described through electrostriction [80], flexoelectricity [81], or phase-transition models [82]. Our findings suggest that $\boldsymbol{E}_{\mathrm{HORZ}}$, inherently generated at the action potential wavefront, may act as the molecular driver for these mechanics. We demonstrate that $\boldsymbol{E}_{\mathrm{HORZ}}$ modulates membrane mechanical properties without measurably changing thickness (Figure 3) and produces spatially distributed optical signatures consistent with vectorial field effects (Figure 4). These responses serve as a mechanistic bridge, connecting the macroscopic mechanical transients observed by Tasaki to discrete, voltage-driven molecular events within the lipid bilayer.

Although we used steady DC $V_{\mathrm{HORZ}}$ to establish a constant $\boldsymbol{E}_{\mathrm{HORZ}}$, the same platform could be extended to AC or pulsed waveforms to enable time-dependent control of $\boldsymbol{E}_{\mathrm{HORZ}}$. Such stimulation would provide experimental paradigms that more closely emulate physiological voltage dynamics and would further allow frequency-dependent modulation of $\boldsymbol{E}_{\mathrm{HORZ}}$, potentially uncovering how $\boldsymbol{E}_{\mathrm{HORZ}}$ interacts with the intrinsic mechanical and electrical resonances of membranes, thereby revealing a new layer of electromechanical signaling.

Our work establishes an experimental route for vector-resolved electric-field actuation in membranes within a model system. While directional control of membrane electric fields is still at an early stage, two parallel challenges lie ahead. The first is the development of effective measurement methods, such as the $\boldsymbol{E}_{\mathrm{HORZ}}$-sensitive dyes suggested by our observations with di-4-ANEPPS, or functional extension of $V_{\mathrm{VERT}}$-responsive fluorescent reporters [83–85], which could visualize three-dimensional electrical landscapes in neuronal networks or epithelial tissues. The second, more formidable challenge is to translate the actuation technology from reconstituted lipid bilayers to living cell membranes. Methods to apply local, controlled $\boldsymbol{E}_{\mathrm{HORZ}}$ directly to cellular membranes will be essential for testing the physiological roles hypothesized here and for connecting fundamental membrane biophysics with *in vivo* function.

For more than seventy years, the voltage-clamp technique developed by Cole [7] and Marmont [9] has been central to membrane biophysics. This powerful tool, which enabled the revolutionary work of Hodgkin and Huxley [6,86], also framed the field conceptually in one dimension: the vertical voltage, $V_{\mathrm{VERT}}$. The platform presented here allows us to move beyond this constraint and begin a systematic exploration of the additional electrical dimensions of the membrane. By providing the first method for generating and interrogating $\boldsymbol{E}_{\mathrm{HORZ}}$, this work lays the foundation for a more complete and physically grounded understanding of the electrical life of the cell.

## 4. Conclusion

This work establishes a bioelectronic platform that enables vector-resolved actuation of membrane electric fields by generating a horizontal electric field ($\boldsymbol{E}_{\mathrm{HORZ}}$) within the hydrophobic core of planar lipid bilayers. By embedding micrometer-scale electrodes in the bilayer torus and achieving teraohm-level insulation under aqueous conditions, the platform delivers $\boldsymbol{E}_{\mathrm{HORZ}}$ actuation while remaining electrically isolated from conventional $V_{\mathrm{VERT}}$ stimulation. Capacitance measurements show that $V_{\mathrm{HORZ}}$ actuation preserves membrane thickness, whereas gramicidin A measurements indicate that $\boldsymbol{E}_{\mathrm{HORZ}}$ modulates membrane mechanical properties. Optical mapping with di-4-ANEPPS further reveals spatially distributed $V_{\mathrm{HORZ}}$ influence across the bilayer and reports vectorial field effects consistent with the summation of $\boldsymbol{E}_{\mathrm{VERT}}$ and $\boldsymbol{E}_{\mathrm{HORZ}}$. As a functional demonstration, $\boldsymbol{E}_{\mathrm{HORZ}}$ selectively and reversibly accelerates KvAP C-type inactivation, establishing field directionality as an orthogonal control axis for membrane protein function. These results provide experimentally tractable access to an in-plane electrical dimension of membranes and support the view that $\boldsymbol{E}_{\mathrm{HORZ}}$ components arising from spatial voltage gradients can operate alongside the canonical $\boldsymbol{E}_{\mathrm{VERT}}$. This platform provides a foundation for developing $\boldsymbol{E}_{\mathrm{HORZ}}$-sensitive readouts and extending directional field actuation toward more complex membrane biointerfaces.

## 5. Experimental Section/Methods

Fabrication and characterization of the $V_{\mathrm{HORZ}}$ electrode chip

The $V_{\mathrm{HORZ}}$ electrode chips were fabricated following the procedure described previously [13,14]. A schematic overview of the fabrication steps is shown in Figure 2D. Teflon films with thicknesses of 12-15 μm (High Sensitivity Membrane Kit, YSI Inc., Yellow Springs, OH, USA) were cut into rectangular sheets of approximately 30 × 40 mm. Small circular apertures (70-180 μm in diameter), across which planar lipid bilayers (PLBs) were subsequently formed, were created by passing an electrical spark generated by an automobile ignition coil. A 200 nm Ti layer was deposited onto the sheet through a metal mask using an electron beam evaporator (VT-43N, ANELVA Corporation, Kyoto, Japan). To electrically isolate the Ti electrodes from electrolyte solutions, the Ti-facing surface of the Teflon sheet was coated with a 300 nm $SiO_2$ layer using either the electron beam evaporator or a sputtering system (QAM-4-S, ULVAC, Chigasaki, Japan). A Pt layer was sputtered onto the exposed Ti regions to prevent the oxidation of the Ti surfaces exposed to the atmosphere. The fabricated devices were washed sequentially with chloroform, ethanol, and toluene, and then immersed in a 2% (v/v) solution of (1H,1H,2H,2H-perfluorooctyl)dimethylchlorosilane (PFDS) in super-dehydrated toluene inside a nitrogen-filled glove box. After incubation at room temperature for 6 h, the devices were rinsed successively with toluene, ethanol, acetone, and chloroform, yielding the $V_{\mathrm{HORZ}}$ electrode chips.

The microstructure around the aperture of the $V_{\mathrm{HORZ}}$ electrode chip was examined using a field emission scanning electron microscope (FE-SEM) (NVision 40, Carl Zeiss, Oberkochen, Germany). Energy-dispersive X-ray spectroscopy (EDS) mapping was performed using a detector (Bruker, Billerica, MA, USA) integrated into the SEM system.

A DC voltage source for the application of $V_{\mathrm{HORZ}}$ was generated by a custom-built circuit comprising dry batteries and a variable resistor [14]. The output voltage range was 0–5 V, and dry batteries were used to minimize mains hum. Leakage currents between the two $V_{\mathrm{HORZ}}$ electrodes, both before and after PLB formation, were measured using a KEITHLEY 2636B SourceMeter (Keithley Instruments, Solon, OH, USA) connected in series with the DC power source and the $V_{\mathrm{HORZ}}$ electrode chip immersed in 0.15 M $Na^+$ buffer (149.2 mM NaCl, 4.7 mM KCl, 2.5 mM $CaCl_2$, 5 mM HEPES-NaOH, [pH 7.3]). The SourceMeter was operated as a high-resolution ammeter with a current resolution of 0.1 fA.

Reagents

The phospholipid used for the KvAP study was 1,2-diphytanoyl-*sn*-glycero-3-phosphocholine (4ME 16:0 PC; DPhPC). The phospholipids used for the other experiments were 1,2-dioleoyl-sn-glycero-3-phosphocholine (DOPC) and cholesterol. DPhPC and DOPC were purchased from Avanti Polar Lipids (Birmingham, AL, USA). Cholesterol was obtained from Fujifilm Wako Pure Chemical (Osaka, Japan) and recrystallized three times from methanol. PFDS was purchased from Gelest Inc. (Morrisville, PA, USA) or Tokyo Chemical

Industry Co. (Tokyo, Japan). All other chemicals were obtained from Nacalai Tesque (Kyoto, Japan) or Fujifilm Wako Pure Chemical.

Cloning, expression, and purification of KvAP

Cloning, expression, and purification of the KvAP channel were performed as described previously [55]. *E. coli* BL21 (DE3) cells carrying the KvAP (14-295 a.a) expression plasmid pET29(a) were cultured in LB medium with kanamycin at 37 °C. When $OD_{600}$ reached 0.6, IPTG (0.2 mM final concentration) was added, and the culture was incubated for 3 h at 37 °C. Cells were collected by centrifugation, and the pellet was resuspended in sonication buffer (20 mM HEPES-KOH [pH 7.4], 200 mM KCl, 0.5 mM PMSF) at 5 mL per g wet cells, followed by sonication. *n*-Decyl-*β*-D-maltoside (DM; 2% (w/v) final concentration; D382, Dojindo Laboratories, Kumamoto, Japan), was added, and the lysate was rotated for 2 h at room temperature. After ultracentrifugation (200,000 g, 30 min, 4 °C), the supernatant was incubated with Talon resin (635652, Takara Bio Inc., Shiga, Japan) (1 mL resin slurry per 20 mL lysate) at 4 °C for 1 h with agitation. The resin was washed with 10 bed volumes of wash buffer (10 mM HEPES-KOH [pH 7.4], 200 mM KCl, 0.25% (w/v) DM, 25 mM imidazole). KvAP was eluted with five bed volumes of elution buffer (10 mM HEPES-KOH [pH 7.4], 200 mM KCl, 0.25% (w/v) DM, 500 mM imidazole). The eluate was concentrated using an Amicon Ultra-4 (50 kDa MWCO) centrifugal filter (UFC805008, Merck Millipore, Burlington, MA, USA) and subjected to size-exclusion chromatography on a Superdex 200 Increase 10/300 GL gel filtration column (28990944, Cytiva, Marlborough, MA, USA) equilibrated with gel filtration buffer (10 mM HEPES-KOH [pH 7.4], 200 mM KCl, 0.25% (w/v) DM).

KvAP-reconstituted liposomes were prepared by dialysis. A mixture containing 0.1 mg $mL^{-1}$ KvAP, 5 mg $mL^{-1}$ DPhPC (850356C), 10 mM HEPES-KOH [pH 7.4], 200 mM KCl, 0.1% (w/v) DM, and 1% (w/v) *n*-octyl-*β*-D-glucoside (25543-14, Nacalai Tesque Inc., Kyoto, Japan) was dialyzed against dialysis buffer (10 mM HEPES-KOH [pH 7.4], 200 mM KCl) using dialysis tubing (132650, Repligen Co., Waltham, MA, USA). The dialysis buffer was replaced five times at 12-h intervals. The final liposome suspension contained 10 mM HEPES-KOH [pH 7.4], 200 mM KCl, 5 mg $mL^{-1}$ DPhPC, and 0.1 mg $mL^{-1}$ KvAP.

Formation of PLBs and recording of vertical membrane currents

Before forming PLBs, the resistance between the two Ti electrodes of each $V_{HORZ}$ electrode chip was verified by connecting them to the input terminals of a patch-clamp amplifier (Axopatch 200B, Molecular Devices, San Jose, CA, USA). Devices that exhibited resistances greater than 250 GΩ in air and 100 GΩ in buffer (0.15 M $K^+$ buffer or 0.15 M $Na^+$ buffer) were used for PLB formation.

PLBs were formed across the aperture in the $V_{HORZ}$ electrode chip using the folding method, as described previously [14]. Briefly, the chip was placed at the center of a Teflon recording chamber, separating the *cis* and *trans* compartments. The area around the aperture on both sides of the chip was pre-coated with a thin layer of n-hexadecane using a cotton swab. Buffer solution (1.4 mL), filtered through a 0.20 µm cellulose acetate filter (Advantec, Tokyo, Japan), was added to each compartment, with the initial water level kept below the aperture. A 30 µL aliquot of a lipid solution was then carefully spread onto the buffer surface in each compartment. After solvent evaporation, a PLB was formed by gradually raising the water level until it surpassed the aperture.

For specific capacitance ($C_{sp}$) measurements, fluorescence imaging, and gramicidin A (gA) recordings, PLBs were formed using a DOPC lipid solution (5 mg $mL^{-1}$; DOPC: cholesterol = 4:1 (w/w)) in chloroform/n-hexane (1:1, v/v). These PLBs were formed in 0.15 M $K^+$ buffer (150 mM KCl and 10 mM HEPES-KOH, [pH 7.4]). For KvAP recordings, PLBs were formed using a 5 mg $mL^{-1}$ DPhPC solution in n-hexane, and currents were recorded in a symmetric recording solution containing 200 mM KCl and 10 mM HEPES-KOH [pH 7.4].

Vertical membrane currents were recorded using an Axopatch 200B patch-clamp amplifier. Signals were filtered and digitized using Digidata 1440A or 1550B interface with pClamp 10.3 or 10.6 software (Molecular Devices). Unless otherwise stated, currents were filtered at 1 kHz (low-pass Bessel filter) and digitized at 10 kHz.

Measurements of specific capacitance ($C_{sp}$)

Simultaneous measurements of vertical membrane currents and bright-field images of PLBs were performed using the setup shown in Figure S1A. A PLB was formed in a homemade Teflon chamber that contained a circular opening for objective-lens insertion and a quartz-glass window on the opposite side for white-light illumination [44]. The quartz glass surface was salinized with PFDS using the same manner as for the $V_{HORZ}$ electrode chip. PLBs were formed in this chamber using 0.15 M $K^+$ buffer and DOPC lipid solution. Because the imaging chamber had a larger surface area than the chamber used for ion-channel recordings, 100 µL of lipid solution was added to each compartment. After PLB formation, a 20 Hz voltage-ramp command (±20 mV) was applied using a function generator (WF1973, NF Corporation, Yokohama, Japan), and the electrical capacitance ($C_{observed}$) was evaluated.

Bright-field images of PLBs were acquired at 10 Hz with a 100 ms exposure time using HCImage Live (Hamamatsu Photonics, Hamamatsu, Japan). The PLB area ($A_m$) was extracted from the images using MATLAB. The PLB boundary was enhanced, binarized, and identified using the “fibermetric”, “imbinarize”, and “bwconncomp” functions, respectively; the “imclose” function was used when boundary correction was required. $A_m$ was then calculated as the area enclosed by the detected boundary.

$C_{sp}$ was evaluated as an index of membrane thickness ($d$) according to:

$$C_{\mathrm{sp}} = \varepsilon_0 \varepsilon_{\mathrm{m}} \frac{A_{\mathrm{m}}}{d} \qquad (1)$$

where $\varepsilon_0$ and $\varepsilon_m$ denote the permittivity of vacuum and relative permittivity of the lipid bilayer, respectively. Because the observed capacitance ($C_{observed}$) comprises both the lipid bilayer capacitance ($C_m$) and stray capacitance ($C_{stray}$), $C_{stray}$ was obtained from the $y$-intercept of the linear fit to the $C_{observed}$-$A_m$ plot, as shown in Figure S1B. $C_{sp}$ was then obtained from:

$$C_{\mathrm{sp}} = \frac{C_{\mathrm{m}}}{A_{\mathrm{m}}} = \frac{C_{\mathrm{observed}} - C_{\mathrm{stray}}}{A_{\mathrm{m}}} \qquad (2)$$

$C_{observed}$ values were averaged over 100 ms and plotted against $A_m$ measured at 10 Hz. Data were excluded for 200 ms before and 300 ms after switching of the DC $V_{HORZ}$ source because current noise during switching prevented reliable estimation of $C_{observed}$. Outliers in $C_{observed}$ caused by electrical noise were removed based on the Smirnov-Grubbs test ($\alpha = 0.05$). Outliers in $A_m$ resulting from the failure of boundary detection were also removed using the same procedure. $C_{sp}$ under $V_{HORZ}$ was calculated as the mean $C_{sp}$ during 4 V $V_{HORZ}$ application over 9.5 s (yellow region in Figure 3A (ii)). $C_{sp}$ without $V_{HORZ}$ was obtained as the mean of $C_{sp}$ during the Off period before and after $V_{HORZ}$ application (gray regions in Figure 3A (ii)), with a combined duration of 9.5 s.

Ratiometric fluorescence imaging

The optical setup shown in Figure 4A was used for ratiometric fluorescence imaging of PLB stained with di-4-ANEPPS (Cayman Chemical, Ann Arbor, MI, USA). PLBs were formed in 0.15 M $K^+$ buffer using DOPC lipid solution, following the same procedure as in the $C_{sp}$ measurements. After PLB formation, 10–30 µL of 1 mg $mL^{-1}$ di-4-ANEPPS stock solution in ethanol was added to the compartment containing the quartz-glass window, yielding a final dye concentration of 3–10 μM. Under these conditions, only one leaflet of the PLB was stained. After 1 h of incubation, the dye-labeled PLB was used for fluorescence imaging.

For ratiometric imaging, the incident light at two excitation bands (centered at 435 nm and 500 nm) produced by a multi-LED light source (pE-4000, CoolLED, Andover, UK) was spectrally narrowed using band-pass filters (transmission bands: 415-455 nm and 488-512 nm). The filtered illumination was further passed through a short-pass filter (cut-off: 550 nm), reflected by a long-pass dichroic mirror (cut-on: 575 nm), and directed onto the PLB through a 63× water-immersion objective lens (NA = 1.0, Carl Zeiss, Oberkochen, Germany). Fluorescence emitted from the PLB was collected through the same objective, transmitted through the dichroic mirror and a long-pass filter (cut-on: 600 nm), and detected by a CMOS digital camera (Orca Fusion, Hamamatsu Photonics, Hamamatsu, Japan). Images were acquired automatically using software (Micro-Manager 2.0), with an exposure time of 100 ms. To minimize photobleaching artifacts, a series of four images was acquired at fixed 300 ms intervals in the following order: 500 nm → 435 nm → 500 nm → 435 nm. For each cycle, $R_{ex}$ was defined as the ratio of the mean intensity of the two 435-nm-excited images to the intensity of the second 500-nm-excited image, with the first 500-nm-excited image discarded (Figure S2).

Determination of the $R_{ex}$-$V_{VERT}$ slope and mapping of the slope ratio

The procedure for determining the $R_{ex}$-$V_{VERT}$ slope is illustrated in Figure S3. A series of nine $R_{ex}$ images was acquired at different $V_{VERT}$ values (−100 to +100 mV in 25 mV increments; 10 s interval) using the method described above. To reduce noise, average pooling (16 × 16) was applied, smoothing the images and decreasing their resolution from 2,304 × 2,304 to 144 × 144 pixels. The resulting pixel size after pooling was 1 × 1 μm. For each pixel, the $R_{ex}$ values obtained at the nine $V_{VERT}$ conditions were used to construct a $R_{ex}$-$V_{VERT}$ plot, which was fitted with a linear function to obtain the slope. The slope ratio was then calculated as the $R_{ex}$-$V_{VERT}$ slope in the presence of $V_{HORZ}$ divided by the slope obtained in its absence for the corresponding pixel. Slope-ratio maps were generated by computing the slope ratio at every pixel. All the data processing and calculations were performed in MATLAB.

Incorporation of gA and analysis of gA channel currents

Gramicidin A (gA; Sigma-Aldrich, St. Louis, MO, USA) was dissolved in methanol and subsequently diluted with 0.15 M $K^+$ buffer. After PLB formation, 10-30 μL of a 10 ng/mL gA solution was added to both the *cis* and *trans* compartments.

The average single-channel lifetime of gA channels was determined as described previously [87], following offline low-pass filtering of the current data at 0.5 kHz. Only channel openings with amplitudes between1.2 to 2.0 pA were included in the analysis. When more than one channel was open simultaneously, a random-number generator was used to assign which channel closed upon the next transition from the open to the closed state. Lifetime distributions were analyzed using Clampfit 11 software (Molecular Devices). The distributions were plotted as log-binned histograms [87] and fitted with single exponential functions: $N(t) = a \exp[\ln(t) - \ln(\tau) - \exp\{\ln(t) - \ln(\tau)\}]$, where $N(t)$ is the number of channels with lifetime $t$, $\tau$ is the mean single-channel lifetime, and $a$ is a scaling factor[87].

Macroscopic current recordings of the KvAP channel

Short (200 ms) and long (3 s) depolarizing $V_{VERT}$ pulses were applied from a *holding potential* of −100 mV. After tail-current recordings at −100 mV, $V_{VERT}$ was stepped to −200 mV to allow recovery from inactivation. Steady-state inactivation was assessed by applying long (15 s) pre-pulses at various holding $V_{VERT}$ values. The extent of inactivation was then evaluated by measuring the peak current elicited by depolarizing pulses to +100 mV.

*G*-$V_{\mathrm{VERT}}$ curve and the steady-state inactivation curve

The *G*-$V_{\mathrm{VERT}}$ curve was obtained by measuring the peak current at depolarized potentials and normalizing these values to the conductance at +40 mV. The steady-state inactivation curve was determined using a three-step protocol. First, channels were fully inactivated by a 3-s depolarization to +100 mV. Second, recovery from inactivation was allowed at hyperpolarized potentials (from −160 to −60 mV) for 15 s. Finally, the peak current elicited by a test pulse to +100 mV was recorded. The peak currents were plotted as a function of the hyperpolarized potentials. Both the *G*-$V_{\mathrm{VERT}}$ and steady-state inactivation relationships are fitted using the Boltzmann function:

$$\frac{a}{1+\exp[-(V_{\mathrm{VERT}}-V_{\mathrm{half}})/k]} \quad (3)$$

where $V_{\mathrm{Half}}$ is the half-activation (or half-inactivation) voltage, $k$ is the slope factor, and $a$ is the scaling constant.

Kinetic analyses

Current traces were ensemble-averaged and fitted using a Hodgkin-Huxley formalism incorporating inactivation [6].

$$I(t) = A\{1-\exp(-t/\tau^{Act})\}^4[h_0-(h_0-h_{\mathrm{inf}})\{1-\exp(-t/\tau^{Inact})\}] \quad (4)$$

where $\tau^{Act}$ and $\tau^{Inact}$are the time constants of activation and inactivation, respectively, and $h_0$ and $h_{\mathrm{inf}}$ denote the initial and steady-state values of the inactivation variable.

Simulation of electric-field distributions using COMSOL Multiphysics

Electric field distributions within the cell membrane were simulated using COMSOL Multiphysics, version 6.2 (COMSOL, Inc., Burlington, MA, USA) employing the *Electrostatics* physics module. The membrane was modelled as a 4-nm-thick dielectric layer with a relative permittivity of 2.2 [1]. For computational efficiency, the lateral dimensions were reduced by a factor of 1000, and the results were rescaled accordingly. Simulations were driven by potential-distance profiles obtained from cortical neurons and dentate granule neurons, as reported previously [15–17]. These profiles were extracted using PlotDigitizer and imported into COMSOL as the intracellular membrane potential via the *Electric Potential* boundary condition. Electric field magnitudes were evaluated at a plane 0.8 nm from the inner membrane surface, and the resulting values are summarized in Table S1.

**Acknowledgements**

We gratefully acknowledge Mr. Kento Abe of the Machine Shop Division, Fundamental Technology Center, Research Institute of Electrical Communication, Tohoku University, for manufacturing specialized equipment used in this study. We also thank Mr. Y. Tsuneta, Mr. R. Yokota, and Dr. K. Kanomata for their valuable experimental assistance.

Declaration of generative AI and AI-assisted technologies in the writing process:

During the preparation of this manuscript, the authors, as nonnative English speakers, used ChatGPT 5 and Gemini 2.5 Flash to improve the language and enhance its readability. After using these tools, the authors reviewed and edited the content as needed and take full responsibility for the final version of the manuscript.

Competing interests:
FH, AH-I, and DT are listed as inventors on a patent application related to the membrane-embedded electrode device used in this study (WO 2020/241453 A1, "Lipid bilayer membrane channel evaluation chip, method for producing same and evaluation apparatus"). All other authors declare no competing interests.

**Data Availability Statement**

All data are available in the main text or the supplementary materials.

Zenodo URL: https://zenodo.org/records/15780293

# Supporting Information

**Genesis of a Horizontal Electric Field within the Lipid Bilayer: A Bilayer-Embedded Actuation Platform**

*Maki Komiya, Madoka Sato, Teng Ma, Hironori Kageyama, Tatsuya Nomoto, Takahisa Maki, Masayuki Iwamoto, Miyu Terashima, Daiki Ando, Takaya Watanabe, Yoshikazu Shimada, Daisuke Tadaki, Hideaki Yamamoto, Yuzuru Tozawa, Ryugo Tero, Albert Martí, Jordi Madrenas, Shigeru Kubota, Fumihiko Hirose, Michio Niwano, Shigetoshi Oiki*, Ayumi Hirano-Iwata**

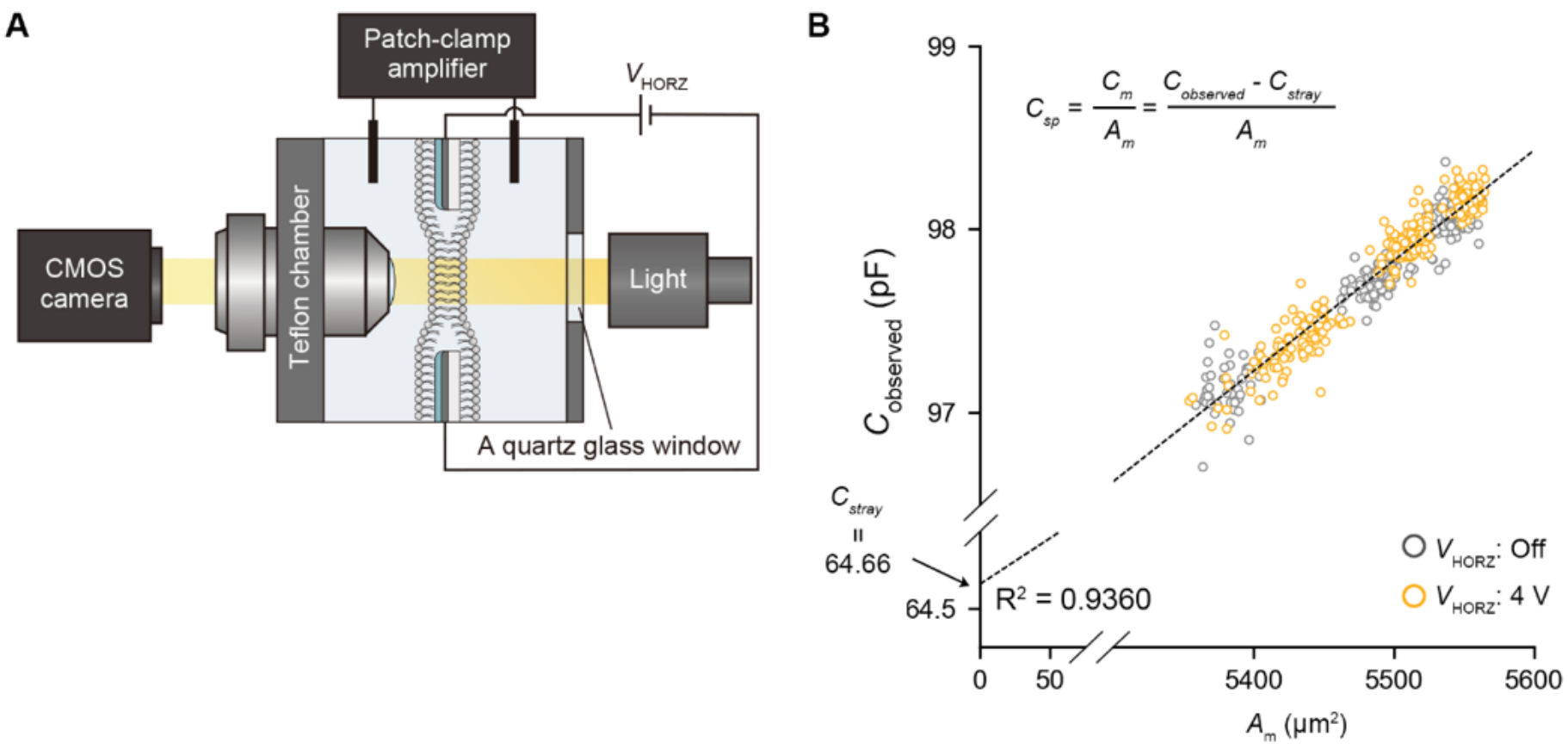

**Figure S1. Evaluation of $C_{sp}$.**
**(A)** Schematic of the experimental setup used for the $C_{sp}$ measurements.
**(B)** Observed capacitance ($C_{observed}$) with and without $V_{HORZ}$ were plotted against the membrane area ($A_m$). Orange, $V_{HORZ}$ = 4 V; Gray, no $V_{HORZ}$. The stray capacitance ($C_{stray}$) was estimated from the y-intercept obtained from the linear regression of the $C_{observed}$-$A_m$ plot. $C_{sp}$ was calculated as ($C_{observed}$ − $C_{stray}$) / $A_m$.

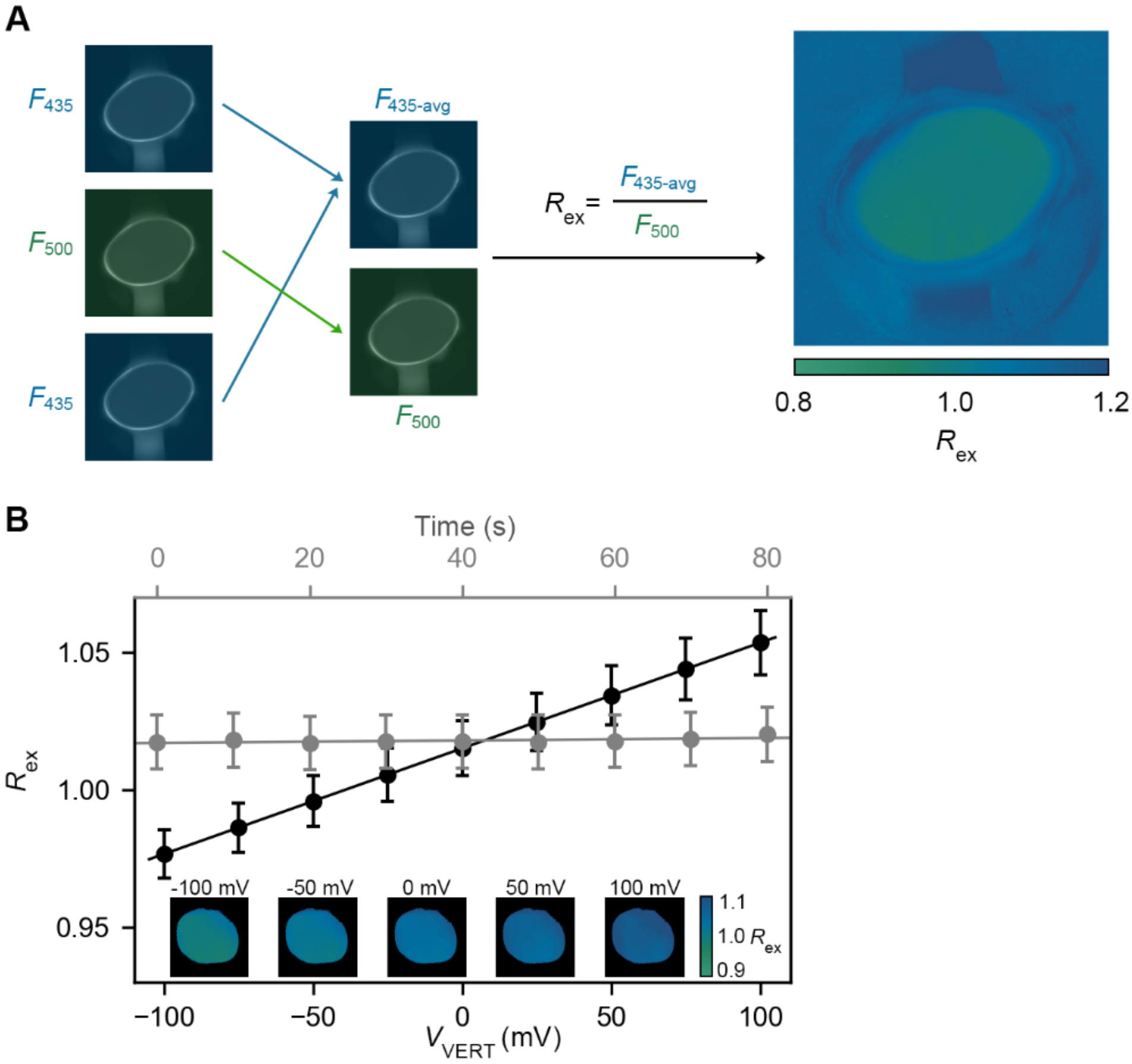

**Figure S2. $R_{ex}$-$V_{VERT}$ relationship of lipid bilayers stained with di-4-ANEPPS.**
**(A)** Acquisition procedure for generating a single $R_{ex}$ image from fluorescence images obtained with 435 and 500 nm excitation.
**(B)** Black circles show the $R_{ex}$-$V_{VERT}$ relationship ($n$ = 13 lipid bilayers), where each point represents the mean $R_{ex}$ value over the entire lipid bilayer region. The slope of the $R_{ex}$-$V_{VERT}$ relation was 3.8 ± 0.2% per 100 mV; the $R_{ex}$ value at $V_{VERT}$ = 0 mV was 1.02 ± 0.01; and the coefficient of determination was $R^2$ = 0.9995 ($n$ = 13 lipid bilayers). For each bilayer, $V_{VERT}$ was sequentially changed from −100 to +100 mV over 80 s. Gray circles show $R_{ex}$ plotted as a function of time over 80 s at $V_{VERT}$ = 0 mV, again using the mean value over the bilayer. The inset shows representative $R_{ex}$ images at $V_{VERT}$ of ±100, ±50, and 0 mV.

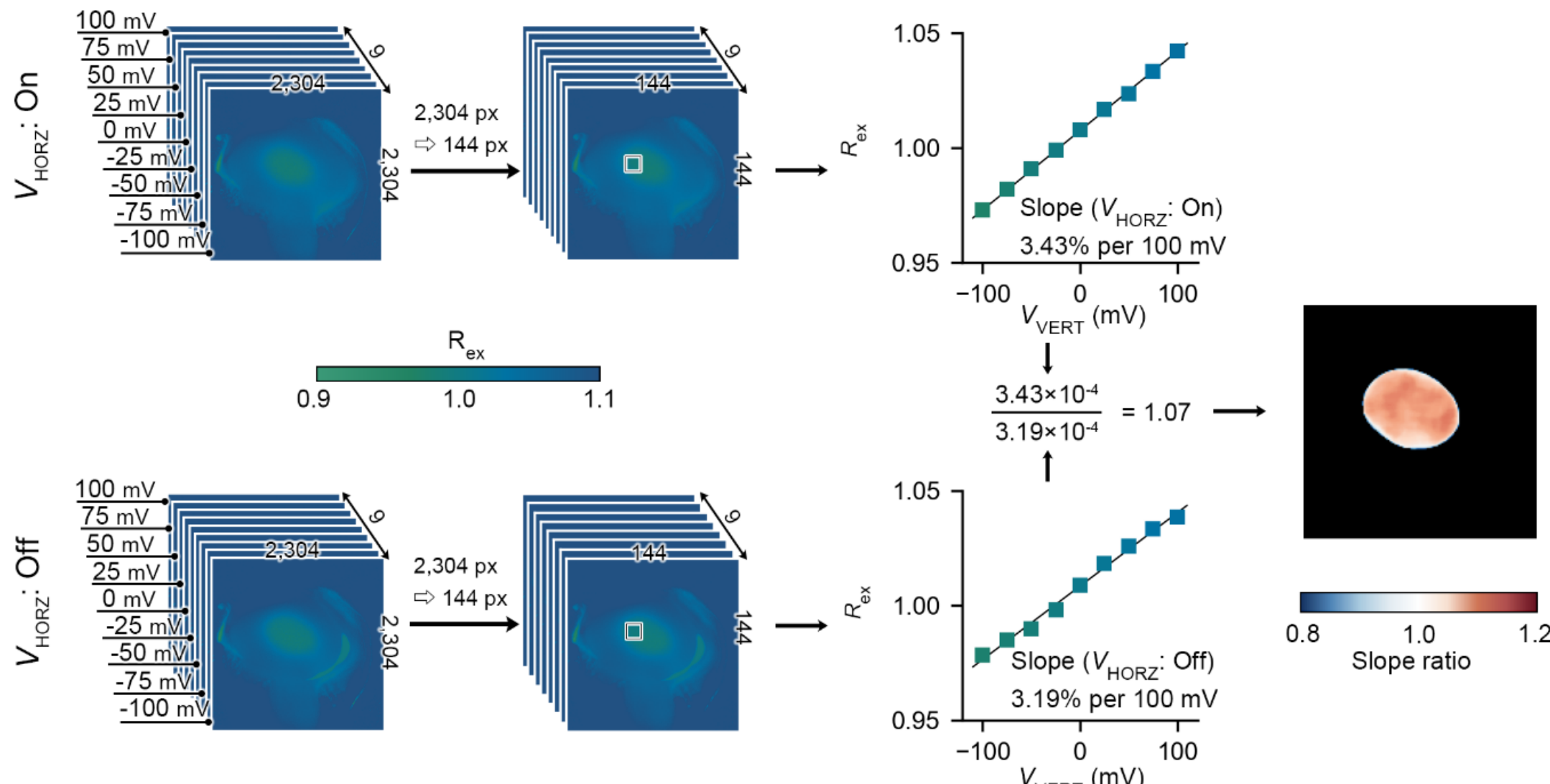

**Figure S3. Procedure for calculating and mapping the $R_{ex}$-$V_{VERT}$ slope ratio.**
A series of nine $R_{ex}$ images acquired at different $V_{VERT}$ values was processed by average pooling (16 × 16) to smooth the images. For each pixel, the corresponding nine $R_{ex}$ values were plotted against $V_{VERT}$, and the resulting $R_{ex}$-$V_{VERT}$ relationship was fitted with a linear function to obtain the slope. The slope ratio was then determined by dividing the slope in the presence of $V_{HORZ}$ by that obtained in its absence for the same pixel. A slope-ratio map was generated by computing this ratio for every pixel. To clearly delineate the lipid-bilayer region, pixels in which the $R_{ex}$-$V_{VERT}$ slope was less than half of the maximum value were colored black in the final map. All the data processing and calculations were performed using MATLAB.

## Supplementary Results

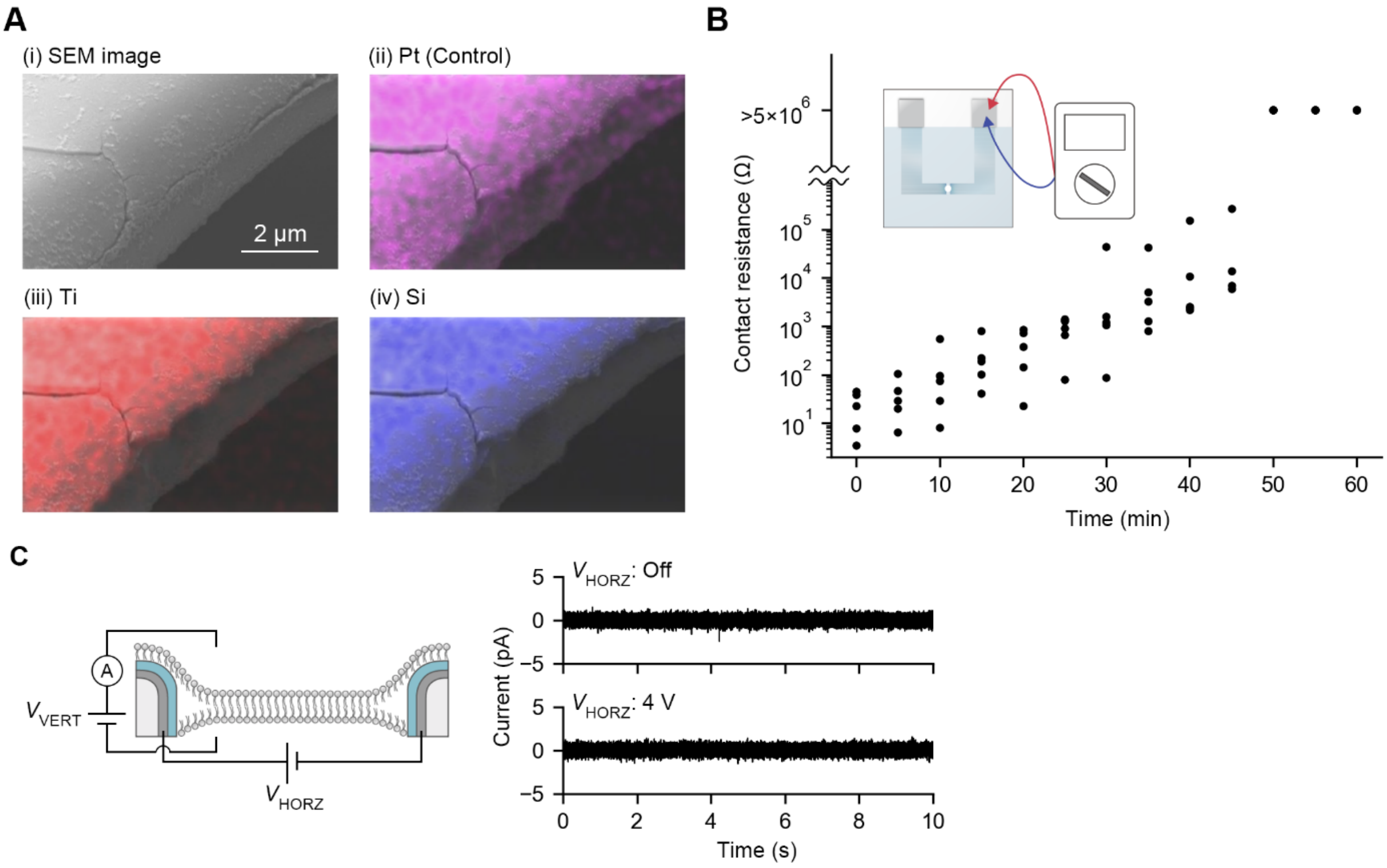

**Figure S4. Characterization of $V_{\mathrm{HORZ}}$ electrode chip.**

**(A)** (i) Scanning electron microscopy (SEM) images and (ii)-(iv) energy-dispersive X-ray spectroscopy (EDS) element maps of the $V_{\mathrm{HORZ}}$ electrode chip around the aperture. A thin Pt coating was applied on both sides of the chip to minimize the charge-up effects and served as a reference in the elemental-mapping experiment. The EDS maps show that the Ti electrodes extended to the mid-height of the aperture wall and that the exposed Ti surfaces are fully covered by the $SiO_2$ insulating layer.

**(B)** Time-dependent changes in the contact resistance, defined as the resistance of the Pt-coated portion of the Ti electrodes, during the application of $V_{\mathrm{HORZ}}$ (4 V) across the two Ti electrodes ($n = 5$).

**(C)** Representative vertical membrane currents recorded at $V_{\mathrm{VERT}} = +100$ mV in the absence and presence of $V_{\mathrm{HORZ}}$ (4 V), illustrating that $V_{\mathrm{HORZ}}$ does not measurably affect vertical leakage.

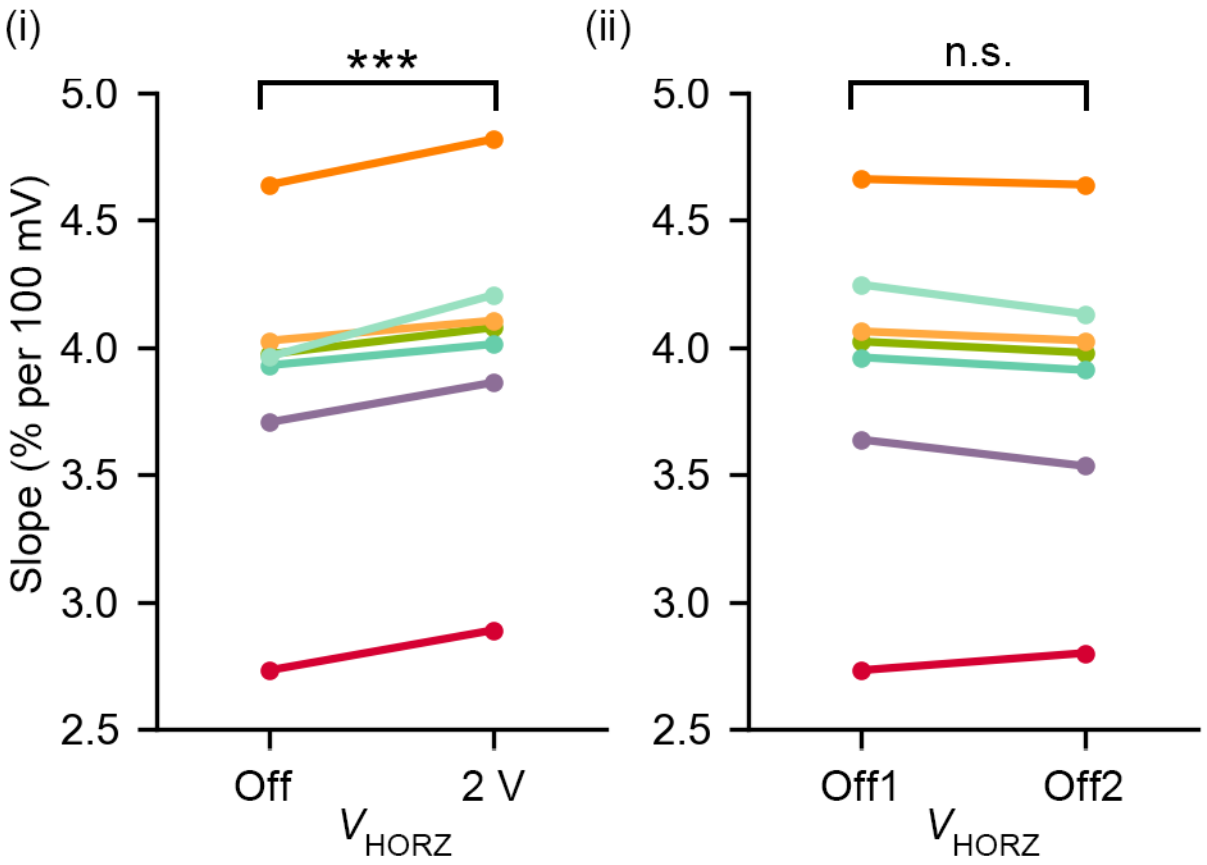

**Figure S5. Effect of $V_{HORZ}$ on the $R_{ex}$-$V_{VERT}$ slope.**
Mean $R_{ex}$-$V_{VERT}$ slope values measured across the entire lipid bilayer ($n$ = 7 lipid bilayers). (i) Slopes with and without the application of $V_{HORZ}$ (2 V). (ii) Two successively measured slopes without $V_{HORZ}$. ***$p < 0.001$ (two-sided paired $t$-test). Data from the same lipid bilayer in panels (i) and (ii) are shown in the same color.

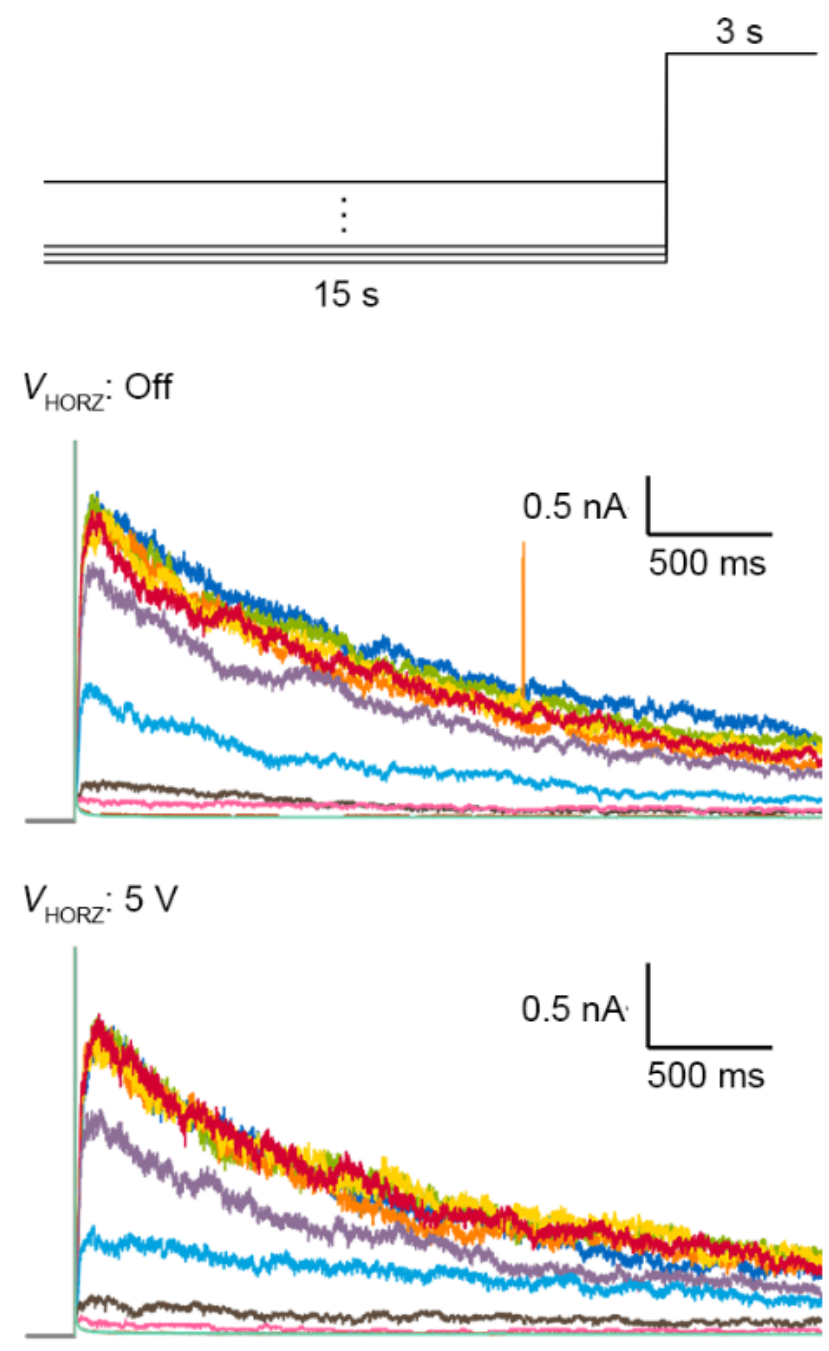

**Figure S6. Steady-state inactivation.**
A depolarizing $V_{VERT}$ pulse of +100 mV for 3 s (top) was applied following a 15-s holding $V_{VERT}$ at various values. The peak current elicited by the depolarizing $V_{VERT}$ pulse decreased progressively as the holding $V_{VERT}$ became less hyperpolarized. A slight acceleration of inactivation was observed under $V_{HORZ}$ application.

**Notes**

**Theoretical Framework of Membrane Elasticity and the Gramicidin A Channel**

The Gramicidin A (gA) channel acts as a sophisticated molecular force probe for lipid bilayers [27,28]. Unlike complex membrane proteins with intricate conformational landscapes, gA forms a transmembrane channel via the reversible, head-to-head dimerization of two monomers [26]. The structural stability and lifetime of this conducting dimer are directly coupled to the mechanical properties of the host lipid bilayer [29,30]. This coupling arises from the hydrophobic mismatch between the channel and the membrane: the hydrophobic length of the gA dimer ($d_0 \approx 2.2$ nm) is typically shorter than the hydrophobic thickness of the unperturbed bilayer ($l$). To shield the hydrophobic residues from the aqueous phase, the lipid bilayer must locally deform (thin) to match the channel boundary.

Energetics of Bilayer Deformation

This deformation incurs an energetic penalty, $\Delta G_{def}$, derived from the continuum elastic properties of the membrane [42,88]. The bilayer resists deformation through two primary modes:

1. **Compression/Expansion**: Resistance to changes in membrane thickness, governed by the area compressibility modulus ($K_a$).
2. **Bending**: Resistance to changes in curvature at the bilayer/solution interface, governed by the bending modulus ($K_c$).

The deformation energy $\Delta G_{def}$ manifests as a **disjoining force** that destabilizes the dimer, promoting dissociation into non-conducting monomers. Consequently, the channel lifetime ($\tau$) becomes an exponential function of the deformation energy: $\tau \propto \exp(-\Delta G_{def} / k_B T)$. A stiffer membrane or a larger hydrophobic mismatch results in a higher $\Delta G_{def}$ and a shorter channel lifetime.

The Nielsen & Andersen Analytical Theory

Nielsen and Andersen established a rigorous continuum elastic model to quantify this interaction [39]. They derived an analytical expression for $\Delta G_{def}$ that decomposes the energy into contributions from the hydrophobic mismatch and the spontaneous curvature of the lipid monolayer ($c_0$). The deformation energy is expressed as:

$$\Delta G_{def}[d_0 - l, c_0] = H_B \cdot (d_0 - l)^2 + H_X \cdot (d_0 - l) \cdot c_0 + H_C \cdot c_0^2 \qquad (5)$$

Deconstruction of the Elastic Coefficients

The equation highlights three critical components of membrane-protein interaction:

- **The Harmonic Term ($H_B \cdot (d_0 - l)^2$)**: This term represents the primary energy cost due to the thickness mismatch. $H_B$ serves as a phenomenological spring coefficient. Crucially, $H_B$ is not a fundamental material constant but a derived function of the bulk moduli ($K_a$ and $K_c$) and the channel radius ($r_0$). It encapsulates the coupled resistance of the bilayer to both compression and the concomitant bending required to match the channel interface.
- **The Cross-Term ($H_X \cdot (d_0 - l) \cdot c_0$)**: This term couples the hydrophobic mismatch to the spontaneous curvature ($c_0$) of the lipid monolayer. It signifies that the energy cost of thinning depends on the lipid's intrinsic tendency to curve. For instance, lipids with negative spontaneous curvature (e.g., PE lipids) can stabilize the concave deformation required for membrane thinning, thereby reducing the energetic penalty via this interaction term.

- **The Curvature Stress Term ($H_C \cdot c_0^2$)**: This term accounts for the energy associated with the spontaneous curvature itself, independent of the mismatch depth, likely arising from the boundary conditions imposed by the channel.

By linking the macroscopic material properties of the lipid bilayer to the microscopic stability of the gA dimer, the Andersen theory provides the essential physical context for interpreting gA channel function. It demonstrates that the membrane is not a passive solvent but an active mechanical continuum that allosterically regulates protein function through the thermodynamic costs of deformation.

**Electrophysiological characteristics of voltage-gated KvAP channels.**

To aid readers who may be unfamiliar with electrophysiology, we provide a brief conceptual overview of voltage-gated channel function using the KvAP channel as an example [55].

1. The channel as a voltage-controlled nanodevice

A voltage-gated ion channel, such as KvAP, can be viewed as a highly evolved, self-regulating nanoscale switch. Its primary function is to open and close an ion-permeable pore in response to changes in the vertical membrane voltage ($V_{\mathrm{VERT}}$), thereby regulating the flow of $K^+$ ions and consequently influencing $V_{\mathrm{VERT}}$ itself.

2. Structure of the KvAP channel

A schematic of the KvAP channel is shown in Figure S7. The channel is composed of a pore domain (PD), which mediates selective $K^+$ ion permeation. The PD contains two gates, the activation gate and the inactivation gate, which operate independently. Surrounding the PD are four voltage-sensor domains (VSDs), each of which responds to changes in $V_{\mathrm{VERT}}$. Each VSD contains a positively charged S4 helix, which moves upward or downward depending on $V_{\mathrm{VERT}}$. Upward movement of the S4 helices is mechanically coupled to the PD and results in opening of the activation gate.

3. The gating cycle: closed, activation, deactivation, inactivation

Voltage-gated channels do not simply switch between “ON” (Open) and “OFF” (Closed). Instead, their behavior is organized by a sequence of distinct conformational states.

- Activation (the primary “ON” switch): When $V_{\mathrm{VERT}}$ changes from a negative value (e.g. −100 mV) to a positive one, the voltage-sensor domains (VSDs; Figure S7, orange) undergo an upward displacement of their positively charged S4 helices (Figure S7, red). After a brief delay, this motion opens the activation gate located at the intracellular end of the pore domain, permitting $K^+$ ions to permeate. This corresponds to the transition from the closed to the open state.
- Cooperative activation (“AND logic circuit”): Opening of the activation gate requires all four VSDs to complete their upward transition. This cooperativity is conceptually analogous to an “AND logic circuit”, giving rise to a steep dependence of channel opening on $V_{\mathrm{VERT}}$: small changes in $V_{\mathrm{VERT}}$ can produce large changes in open probability.
- Deactivation (The primary “OFF” switch): When the membrane potential returns to a negative value, the activation gate closes rapidly. This corresponds to the transition from the open back to the closed state.
- Inactivation (The “Auto-OFF” or “Circuit Breaker” function): If depolarization is sustained for a longer period (typically seconds), a separate inactivation gate within the pore domain closes, eliminating ionic conduction. This transition from the open state to the inactivated state occurs even though the depolarizing stimulus is still present. This mechanism is functionally analogous to that of a self-resetting circuit breaker that prevents continuous current flow under prolonged stimulation.

- Recovery from Inactivation: To regain function, the inactivated channel must first return to the closed state. This requires the membrane potential to be held at a sufficiently negative voltage for a comparatively long period (seconds). Only after this recovery can the channel reopen upon renewed depolarization.

4. Experimental characterization: input ($V_{\mathrm{VERT}}$ protocol) and output (current)

To characterize these gating properties, we used a voltage-clamp amplifier to apply defined $V_{\mathrm{VERT}}$ steps (input) and to record the resulting ionic currents (output).

- Holding Potential: The baseline negative $V_{\mathrm{VERT}}$ (e.g., −100 mV) at which channels are maintained in the closed, ready-to-activate state.
- Depolarizing Pulse: A step to a positive $V_{\mathrm{VERT}}$ that triggers activation and, if maintained for a sufficiently long duration, initiates inactivation.
- Macroscopic Current: Because thousands of channels are present in the membrane, the recorded current reflects the summed activity of all channels. The time course of this current trace reveals the kinetics of the underlying gating processes. For example, the initial sigmoidal corresponds to activation, whereas the subsequent slow decay during a prolonged depolarizing pulse reflects entry into the inactivated state.

5. Key parameters for quantification

The behavior of the channel is quantified using several key parameters derived from these $V_{\mathrm{VERT}}$-clamp experiments:

- Conductance-$V_{\mathrm{VERT}}$ ($G$-$V$) curve (Figure 5C): This describes steady-state activation. It shows how the channel's open probability depends on $V_{\mathrm{VERT}}$, thereby defining the $V_{\mathrm{VERT}}$ range over which the channel activates.
- Steady-state inactivation curve (Figure 5F): This describes the $V_{\mathrm{VERT}}$ dependence of recovery from inactivation. It plots the fraction of channels that remain available for activation after being held at different negative $V_{\mathrm{VERT}}$ values, thereby defining the $V_{\mathrm{VERT}}$ range required to reset the "circuit breaker."
- Time constants ($\tau$) (Figure 5E): These parameters describe the kinetics of gating transitions. $\tau^{\mathrm{Act}}$ reflects how rapidly the activation gate opens, whereas $\tau^{\mathrm{Inact}}$ reflects how rapidly the inactivation gate closes.

This framework provides the necessary background to understand why the selective acceleration of inactivation kinetics ($\tau^{\mathrm{Inact}}$) by $\boldsymbol{E}_{\mathrm{HORZ}}$, while leaving the primary activation properties ($G$-$V_{\mathrm{VERT}}$ curve) unaffected, is a striking and physiologically significant observation.

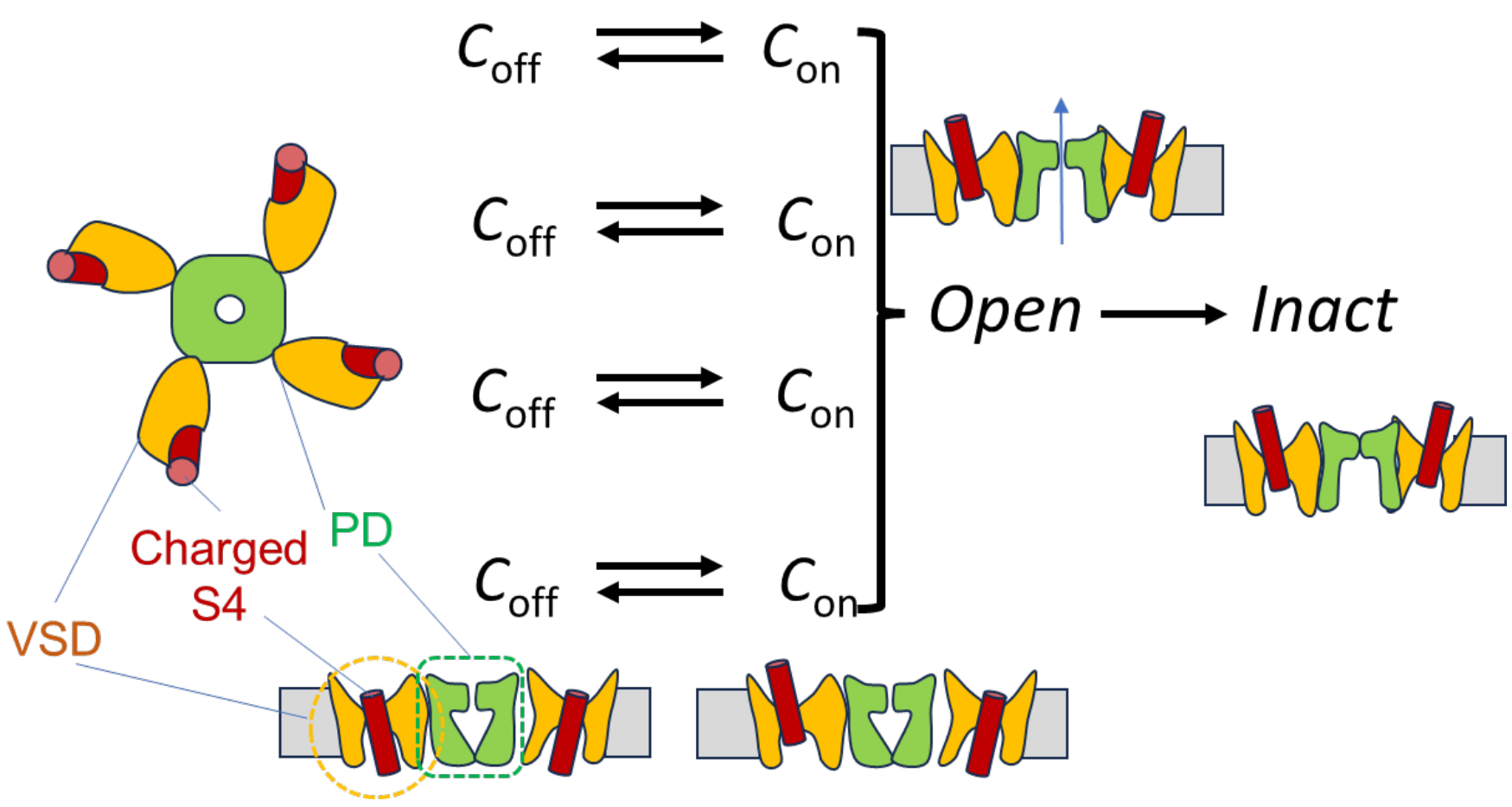

**Figure S7. Gating states of voltage-gated potassium channels.** A voltage-gated channel can be conceptualized as a $V_{\mathrm{VERT}}$-controlled switch that cycles between three principal functional states: (C) Closed, (O) Open, and (I) Inactivated.

Left: Schematic representation of the KvAP channel. The channel comprises a pore domain (PD), which forms the ion-conduction pathway, and four surrounding voltage-sensor domains (VSDs), which detect changes in $V_{\mathrm{VERT}}$. The PD contains two sequential gates: an activation gate located on the intracellular side (lower) and an inactivation gate on the extracellular side (upper). In the Closed state, the activation gate is shut while the inactivation gate remains open. Upon depolarization, upward movement of the VSD S4 helices opens the activation gate, permitting ion flow because the inactivation gate is initially open. During sustained depolarization, the inactivation gate closes slowly; once closed, it remains shut as long as $V_{\mathrm{VERT}}$ is held at a positive value.

Right: Simplified state diagram of gating process. A depolarizing $V_{\mathrm{VERT}}$ step drives the transition from the Closed state to the ion-conducting Open state (activation), a process requiring all four VSDs to switch to their activated conformation. Prolonged depolarization then drives the transitions to the non-conducting Inactivated state. Recovery from inactivation requires repolarization to sufficiently negative $V_{\mathrm{VERT}}$ values for an appropriate duration, after which the channel returns to the Closed, activatable state.

**Table S1.**
**Estimated maximum horizontal electric field in cell membranes obtained from COMSOL simulations.**
Intramembrane electric-field distributions were calculated using COMSOL Multiphysics on the basis of previously reported vertical potential-distance profiles.

| References | Cell type | Myelination | Calculated maximum horizontal electric field (V $m^{-1}$) |
|---|---|---|---|
| [15] | Dentate granule neuron | Unmyelinated neuron | $9.2 \times 10^2$ |
| [16] | Cortical pyramidal neuron | Unmyelinated region (axon initial segment) | $2.4 \times 10^3$ |
| [17] | Cortical pyramidal neuron | Unmyelinated region (axon initial segment) | $2.4 \times 10^3$ |

## Appendix

### Membrane electric field during action potential propagation

Consider a cylindrical structure enclosed by a cell membrane, along which an action potential propagates in the longitudinal ($x$) direction (Figure S8). The $y$- and $z$-axes are oriented perpendicular to the $x$-axis and to each other, as shown in Figure S8. The electric field vector $\boldsymbol{E}(\boldsymbol{r})$ at position $\boldsymbol{r} = (x, y, z)$ can be expressed as the gradient of electrostatic potential $\phi(\boldsymbol{r})$, as follows:

$$\boldsymbol{E}(\boldsymbol{r}) = -\nabla\phi(\boldsymbol{r}) = -\left(\hat{x}\frac{\partial\phi}{\partial x},\ \hat{y}\frac{\partial\phi}{\partial y},\ \hat{z}\frac{\partial\phi}{\partial z}\right). \quad (6)$$

where $\hat{x}$, $\hat{y}$, and $\hat{z}$ are unit vectors along the respective axes. If the action-potential wavefront propagates parallel to the $yz$-plane, then $\frac{\partial\phi}{\partial y} = 0$. Under this condition, Equation (6) simplifies to:

$$\boldsymbol{E}(\boldsymbol{r}) = -\nabla\phi(\boldsymbol{r}) = -\left(\hat{x}\frac{\partial\phi}{\partial x},\ \hat{z}\frac{\partial\phi}{\partial z}\right) = \boldsymbol{E}_{\mathrm{HORZ}} + \boldsymbol{E}_{\mathrm{VERT}}. \quad (7)$$

Thus, only the $xz$-plane needs to be considered in the subsequent discussion (see Figure 1). In the resting state, the intracellular potential is spatially uniform from $x$ to $x+\Delta x$, equal to the resting potential. In this case, $\frac{\partial\phi}{\partial x} = 0$, and the electric field becomes:

$$\boldsymbol{E}(\boldsymbol{r}) = -\hat{z}\frac{\partial\phi}{\partial z} = \boldsymbol{E}_{\mathrm{VERT}}. \quad (8)$$

**Figure S8.** Definition of $x$-, $y$-, and $z$-axes on the axonal cell membrane.

A similar situation holds in the excited region: although the polarity of $\boldsymbol{E}_{\mathrm{VERT}}$ reverses relative to the resting region, $\frac{\partial\phi}{\partial x} = 0$ and the field remains purely vertical.

In the transition zone between resting and excited regions, the electric field $\boldsymbol{E}(\boldsymbol{r})$ contains both an $x$-component ($\boldsymbol{E}_{\mathrm{HORZ}}$) and a $z$-component ($\boldsymbol{E}_{\mathrm{VERT}}$). Figure S9 schematically illustrates the spatial distribution of the electrostatic potential $\phi(\boldsymbol{r})$ in this intermediate region. In the figure, black lines indicate equipotential contours, and arrows represent the direction and magnitude of $\boldsymbol{E}(\boldsymbol{r})$. The horizontal component $\boldsymbol{E}_{\mathrm{HORZ}}$ reaches its maximum at positions where the spacing between the equipotential lines along the $x$-axis is narrowest. This condition occurs in the immediate vicinity of the point where the intracellular potential reverses its polarity.

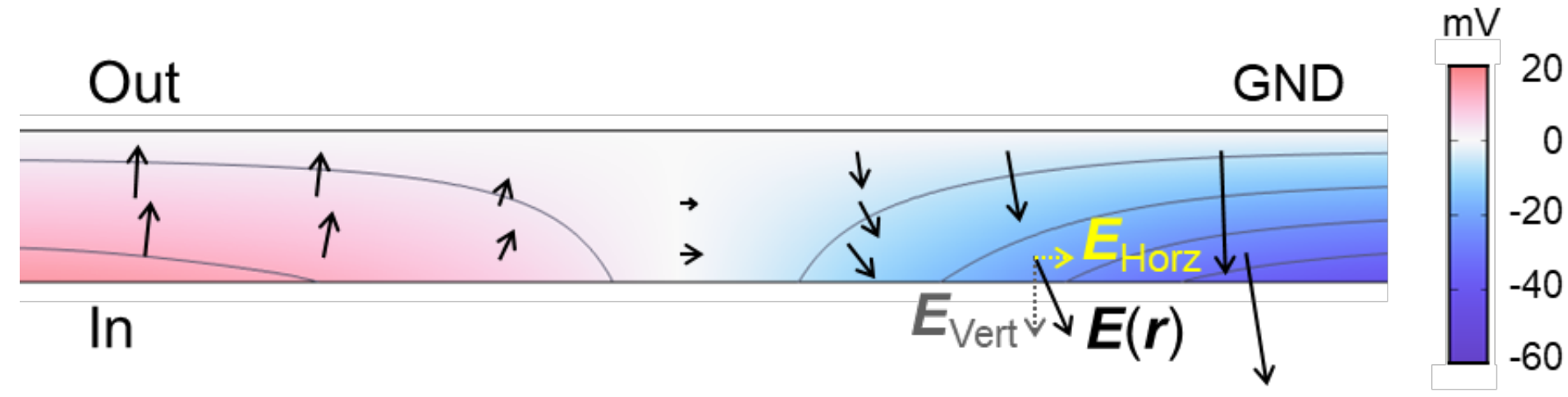

**Figure S9.** Schematic illustration of the electrostatic potential $\phi(\boldsymbol{r})$ within the cell membrane in the intermediate region between the resting state (blue) and excited state (red). Equipotential lines (black) and the local electric field vectors $\boldsymbol{E}(\boldsymbol{r})$ (arrows) are shown to highlight the coexistence of vertical ($\boldsymbol{E}_{\mathrm{VERT}}$) and horizontal ($\boldsymbol{E}_{\mathrm{HORZ}}$) field components.